\documentclass[twocolumn,showpacs,amsmath,amssymb,pra,superscriptaddress]{revtex4-2}

\usepackage{graphicx}
\usepackage{dcolumn}
\usepackage{bm}
\usepackage{subfigure}
\usepackage{blindtext}

\begin{document}


\title {
Accounting all contributions for the Van Vleck paramagnetism and the Langevin diamagnetism
from first principles: application to diamond
}

\author {A. V. Nikolaev}

\affiliation{
Skobeltsyn Institute of Nuclear Physics, Moscow
State University, Vorob'evy Gory 1/2, 119234, Moscow, Russia
}

\author {I. I. Vlasov}

\affiliation{Prokhorov General Physics Institute of the
Russian Academy of Sciences, 119991 Moscow, Russia}

\author {L. L. Tao}

\affiliation{School of Physics, Harbin Institute of Technology, Harbin 150001, China}


\begin{abstract}
A general method for calculating magnetic susceptibility ($\chi$) in dielectrics within a single choice of magnetic gauge for the whole crystal is presented.
On the basis of the method, accounting for all contributions to the Van Vleck paramagnetism
and Langevin (Larmore) diamagnetism, a full-scale {\it ab initio} calculation of $\chi$ in diamond is performed.
Unfamiliar contributions to $\chi$ includes a Van Vleck contribution from the interstitial region and an offset contribution from the muffin-tin (MT) sphere,
appearing due to the change of the MT-sphere magnetic moment when the sphere is displaced from the origin.
Although the Langevin diamagnetism explicitly depends on the choice of the origin,
its sum with the Van Vleck term remains invariant, which is demonstrated on the basis of the gauge invariance of
the magnetic vector potential.
The derived expressions have been applied to {\it ab initio} calculations of magnetic susceptibility of
the crystalline diamond within the linear augmented plane wave method (LAPW).
With the diamond unit cell having the inversion symmetry, the magnetic (Van Vleck) calculations require the irreducible part of the Brillouin zone accounting for half of the whole zone, i.e. 24 times larger than that in the absence of magnetic field.
Investigating possible anisotropy of $\chi$, we
calculate it for 74 different directions of $H$ (belonging to Lebedev surface grid points),
and demonstrate that the actual value of $\chi$ remain isotropic.
The obtained volume magnetic susceptibility in diamond lies in the range $-16.27-16.72$ (with the Langevin contribution $-39.22-39.94$ and the Van Vleck contribution $-22.94-23.22$), in units $10^{-7}$, which compares well with the experimental data and other calculations.
%
\end{abstract}


\maketitle

\section{Introduction}
\label{sec:int}

The task of predicting magnetic properties of solids and nano-structures is of fundamental importance.
If, thanks to the rapid development of {\it ab initio} calculations, which are now regularly used in experimental and theoretical studies, the problem of obtaining and predicting the electronic structure can be considered generally solved, then the situation with magnetic properties is still far from this stage.
On the other hand, the progress in practical magnetic applications is closely related to spintronics, which manipulates very sensitive electronic or
nuclear spin degrees of freedom \cite{spin}.
In particular, much attention recently has been drawn by new research in detection, characterization and manipulation by
nitrogen-vacancy (NV$^-$) centers \cite{Barry,diam-22} and $^{13}$C nuclear spins in diamond \cite{King2,King,C13-rev}.
The problem is of great practical importance -- it is believed that the emerging techniques
can be used in magnetometer of extreme sensitivity \cite{Barry,Mul} and nano-scale resolution \cite{Bala,Maze,Zhao}.
Other applications include bioimaging under ambient conditions \cite{Bala,McGui} and quantum information processing \cite{Buck,Pez,Maze}.
Since correlated $^{13}$C nuclear spin states can be very long-lived \cite{Stev},
nuclear spin hyperpolarization generated from NV-vacancy centres in diamond is considered as a possible platform for polarization
transfer to samples for nuclear magnetic resonance spectroscopy and magnetic resonance imaging \cite{C13-rev,King,Shagi,King2,Green,Pagli,Niz}.
To understand the mechanism and advance the NV$^-$ and $^{13}$C based technologies, knowledge
of the host material magnetic properties (both diamagnetic and
paramagnetic) is undoubtedly necessary.
There are very few experimental works on magnetic behavior of diamond \cite{Hud,Her} in the literature, which is
undoubtedly related to the technical complexity of such measurements \cite{Nik-book}.
Therefore, the study of the magnetic response of diamond at {\it ab initio} level,
which is free from assumptions and suggestions, is of primary interest.

Early attempts to calculate the magnetic susceptibility of diamond and other insulators were based on partially semiempirical models \cite{Hud,Sukh75,Chadi75,Nik-VV}.
The first method for the calculation of the magnetic susceptibility of insulators from first principles was developed in 1996 in the pioneer work
of Mauri and Louie \cite{ML96}. The external magnetic field $H$ in this work was modulated with a finite wave vector $\vec{q} \neq 0$,
and the diamagnetic part (proportional to $A^2 \propto q^{-2}$, where $A$ is the vector potential) by means of the $f$-sum rule was rewritten through
a sum $g(\vec{k},\vec{k})$, which was also
used for the calculation of the paramagnetic part ($ \propto \vec{p} \vec{A}$).
Thus, both contributions are expressed in terms of the same function $g(\vec{k},\vec{k}')$, which in practice removes the numerical instability
associated with the $q^{-2}$ dependence in $\chi$.
The full response was calculated taking the limit at $\vec{q} \rightarrow 0$, i.e. at the infinite wavelength of the modulation.
The $f$-sum rule implies a certain choice of gauge.
In \cite{Gregor99} it is shown that for finite systems the Mauri, Pfrommer, and Louie (MPL) method \cite{MPL},
which extends the concept of \cite{ML96} to the problem of chemical shift calculations,
is equivalent to the continuous ($d(r)=r$) set of gauge transformation (CSGT).
In the following the focus of the research was shifted to {\it ab initio} calculations of nuclear magnetic resonance (NMR) chemical shifts on the basis of
the MPL approach \cite{MPL}. The early version of the method \cite{ML96,MPL} was limited to light elements with hard pseudopotentials.
In later development these limitations have been overcome in subsequent works \cite{Pickard01} and \cite{Yates07}.

Following the approach of \cite{Pickard01,Yates07}, the formalism was developed and implemented for a full potential all electron augmented plane wave (APW) method
used for precise calculations of electron band structure \cite{Laskow12,Laskow14}.
In this method there are no restrictions imposed on the electron density or induced electron current, and the integration is performed without further approximations.
Nevertheless, the formalism still requires the modulation with the wave vector $\vec{q}$ of either the magnetic field \cite{ML96,MPL} or the position operator \cite{Pickard01},
and as before the actual results are obtained in the limit $q \rightarrow 0$.
As \cite{ML96,MPL} the method employs the continuous ($d(r)=r$) gauge for the valence electrons.
Although the studies \cite{Laskow12,Laskow14} aim mainly the NMR chemical shifts, the magnetic susceptibility can also be calculated.
The obtained values of $\chi$ for some fluorides, oxides, chlorides and bromides are quoted in Table~I of Ref.~\cite{Laskow14}.

In the present study we work within a single choice of the gauge for the vector potential $\vec{A}$, Eq.\ (\ref{i1a}), with its origin, $\vec{R}_0$, situated at the center
of the first primitive unit cell ($n=1$), Fig.\ \ref{fig1}, located at the origin of the coordinate system. This, in particular, implies, that
the vector potential $\vec{A}$ is different in all other unit cells ($n \neq 1$). However, as discussed in Sec.~\ref{sec:extra} and Appendix~A below,
the contributions from other cells will be the same due to the gauge invariance.
Our approach differs from the continuous gauge ($d(r)=r$) used in other treatments \cite{ML96,MPL,Laskow12,Laskow14}.
We also do not use the modulation of the external magnetic field with the wave vector $q$, working in the fully static environment, $q \equiv 0$.
Our method for the Van Vleck and Langevin contributions \cite{VV} is directly applicable to any solid, not only to dielectrics, but also to metals.
Of course, in the case of metals,
the Pauli paramagnetism and the Landau diamagnetism should be taken into account as well \cite{AM,Lan1}.

Our main goal here is to study possible anisotropy of $\chi$ in cubic diamond. It is worth noting that the magnetic anisotropy does
present in cubic crystals -- e.g. the well known textbook example of magnetocrystalline anisotropy in ferromagnetic cubic structures of iron and nickel \cite{Kit}.
On the other hand, theoretically it is shown that there is an explicit dependence of $\chi$ on the direction of $\vec{H}$ for cubic alkali metals
due to the Landau diamagnetism \cite{Nik1}.
Therefore, {\it a priory} we can not state that in the case of diamond there is no such an anisotropy.

In diamond and other dielectrics the magnetic susceptibility is limited only by Van Vleck and Langevin contributions.
Although these two terms are known for many decades, in the next section
we will see that within the single choice of gauge for the whole crystal, adopted in the present study, they can take unfamiliar forms,
such as those originating from the interstitial region (Sec.\ \ref{sub:VVI} below) or resulting in offset terms
related to the displacement of MT-spheres (Sec.\ \ref{sub:MT} and Eq.\ (\ref{MT3}) below).
Finally, we note that all our numerical calculations are based on {\it ab initio} band structure calculations of diamond,
using the full potential linear augmented plane wave method (FLAPW) \cite{blapw,wien2k}
as implemented in Ref.\ \cite{Nik2}, with technical parameters given in Sec.\ \ref{sub:basis}.

The paper is organized as follows. In Sec.~\ref{sec:extra} we discuss the origin of additional terms in our single gauge approach,
in Sec.~\ref{sec:VV} and Sec.~\ref{sec:LL} we derive explicit expressions for the Van Vleck paramagnetic and Langevin diamagnetic contributions.
In Sec.~\ref{sec:diamond} the method is applied to the diamond, where we calculate $\chi$ for various directions of $\vec{H}$ and
conclude that $\chi$ is isotropic. Our conclusions are summarized in Sec.~\ref{sec:con}.

\section{Method: extra contributions to $\chi$ in the single gauge approach}
\label{sec:extra}

\begin{figure}
\resizebox{0.45\textwidth}{!} {
\includegraphics{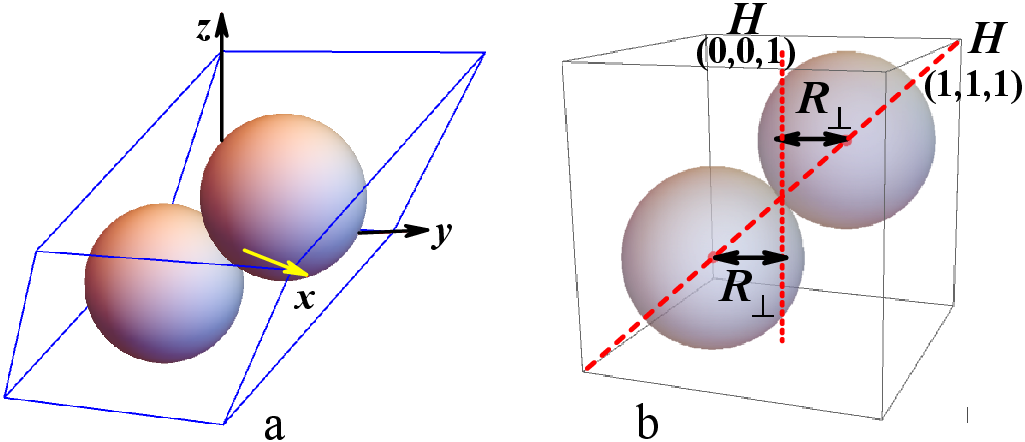}
}

\caption{
(a) The diamond unit cell with two touching MT-spheres with the centers at $\vec{R}_c=\pm(1,1,1)\,a/8$.
The origin $O\; (0,0,0)$ is at the center of the unit cell, which is also the inversion center.
(b) Perpendicular components $R_{\perp}$ of the MT-sphere centers $\vec{R}_c$ in respect to two lines of zero vector potential $\vec{A}$
(for $\vec{H} \parallel (1,1,1)$ and $\vec{H} \parallel (0,0,1)$)
} \label{fig1}
\end{figure}

As mentioned in the introduction, we work with a single choice of magnetic gauge throughout the whole crystal.
In the final expressions given in Sec.~\ref{sec:VV} and Sec.~\ref{sec:LL} for valence electrons the gauge origin $\vec{R}_0$ [see Eq.~(\ref{i1a})] coincides
with the origin of the coordinate system, i.e. $\vec{R}_0=0$, and the focus is on the first ($n=1$) unit cell, Fig.~\ref{fig1}.
However, in this section for methodological purposes we also consider the case $\vec{R}_0 \neq 0$,
and discuss the most important consequences of our approach: the dependence of partial contributions on the choice of the gauge origin,
and the appearance of additional contributions to $\chi$.
Since we aim to study possible anisotropy of $\chi$, in all equations below we stress out the dependence on the direction of~$H$.

In general, there are two different mechanisms for diamagnetism: the Landau diamagnetism of itinerant electrons \cite{Lan1}
and the Langevin diamagnetism \cite{VV} (equivalent to the Larmor precession in the presence of the center of attraction).
In dielectrics or semiconductors the electron spectrum exhibits a gap at the Fermi level,
separating the valence and conduction band. The density of states at the Fermi level (or chemical potential) $N(E_F) = 0$,
which implies that the Pauli paramagnetism is excluded, $\chi_{Pauli}=0$.
In such a case, the Landau diamagnetism is also absent, because
a fully occupied electron band results in zero contribution, $\chi_{Landau}=0$ \cite{Nik1}.

Thus, the diamagnetism in our case can be exclusively of the Langevin ($\chi^{dia}$) type and the paramagnetism of the Van Vleck ($\chi^{para}$) type.
The Langevin diamagnetism $\chi^{dia}$ includes the core and valence electron contributions ($\chi^{dia}_c$ and $\chi^{dia}_{val}$) given by \cite{Lan1,Lan}
\begin{eqnarray}
   \chi^{dia} = -\frac{e^2}{4mc^2} \langle r_{\perp}^2 \rangle ,
\label{i1}
\end{eqnarray}
where $\langle r_{\perp}^2 \rangle$ stands for averaging over the core electron density $\rho_{core}(r)$ or valence electron density $\rho_{val}(r)$.
Here $r_{\perp}$ stands for the component of the radius vector $\vec{r}$, perpendicular to the direction of the magnetic field $H$, $\vec{n}_H=\vec{H}/H$.
Equivalently, the component $\vec{r}_{\perp}$ is perpendicular to the zero line of the vector potential $\vec{A}$, because
\begin{eqnarray}
   \vec{A} = \frac{1}{2}\, \vec{H} \times (\vec{r} - \vec{R}_0) = \frac{1}{2}\, r_{\perp} H\, \vec{e}_{\perp, H}
\label{i1a}
\end{eqnarray}
is perpendicular to $\vec{r}_{\perp}$ and to the direction of $\vec{n}_H$,
with the unit vector $\vec{e}_{\perp, H} \sim \vec{r}_{\perp} \times \vec{n}_H$.
The vector $\vec{R}_0$ corresponds to the gauge origin. We start with $\vec{R}_0 = 0$, but later consider cases with $\vec{R}_0 \neq 0$.
For the averaged value, $\langle r_{\perp}^2 \rangle$, we have,
\begin{eqnarray}
   \langle r_{\perp}^2 \rangle = \int_{V} \rho(r)\, r_{\perp}^2 \; dv ,
\label{i1b}
\end{eqnarray}
where the integration is taken over the unit cell region (with the volume $V$).
In the case of core electron shells the electron density $\rho_c$ is spherical and confined by the interior region of MT-spheres
centered at nuclei. In that case the region of integration in Eq.~(\ref{i1b}) is simply the MT-spheres, shown for the diamond unit cell in Fig.\ \ref{fig1}.
For the valence electrons ($\rho_{val}$), the integration should be performed over the whole unit cell including the interstitial region.
%
%

The important feature of the Langevin diamagnetism $\chi^{dia}$ is that it explicitly depends on the choice of the gauge origin through $\vec{R}_0$ in Eq.\ (\ref{i1a}).
Suppose that we find $\chi^{dia}_{MT,0}=-C_L \langle r_{\perp}^2 \rangle_c$ (here $C_L=e^2/4mc^2$) for the electrons confined strictly inside a MT sphere,
whose center $\vec{R}_c$ coincides with the origin
(i.e $\vec{R}_c = \vec{R}_0 = 0$).
If now we calculate $\chi^{dia}_{MT,R}$ for the MT-sphere displaced from $O(0,0,0)$ through the translation by the vector $\vec{R}$,
we obtain
\begin{eqnarray}
   \chi^{dia}_{MT,R} = -C_L \langle (\vec{r} + \vec{R})_{\perp}^2 \rangle_c  \nonumber \\
   = \chi^{dia}_{MT,0} - C_L R_{\perp}^2 |Q^{MT}| ,
\label{i2}
\end{eqnarray}
where $R_{\perp}$ is the component of $\vec{R}$ perpendicular to $\vec{H}$,
and $Q^{MT}$ is the total electron charge inside the MT-sphere. In Eq.\ (\ref{i2}) we have assumed that the dipole momentum
of the whole sphere, counted from its center, is zero,
i.e. $\langle r_{\perp} \rangle_c = 0$. Note, that $\langle r_{\perp} \rangle_c \neq 0$ can occur in ferroelectrics; in most cases,
including diamond, it is absent. $R_{\perp}$ for two spheres in the diamond unit cell explicitly depends on the direction of $H$.
For example, as shown in Fig.\ \ref{fig1}b for the field direction [1,1,1] $R_{\perp} = 0$, whereas for [0,0,1] $R_{\perp} \neq 0$.
From Eq.\ (\ref{i2}) it follows that $|\chi^{dia}_{MT,R}| > |\chi^{dia}_{MT,0}|$.

The total susceptibility, $\chi^{tot}=\chi^{dia}+\chi^{para}$, however, includes also the paramagnetic Van Vleck term \cite{Lan},
\begin{eqnarray}
   \chi^{para} = 2 {\sum_{i' \neq i} }' \frac{|(M_H)_{i'i}|^2}{E_{i'}^{(0)}-E_{i}^{(0)}} .
\label{i3}
\end{eqnarray}
The sum of two terms, $\chi^{dia}+\chi^{para}$, does not depend on the choice of the gauge origin [$\vec{R}_0$ in Eq.\ (\ref{i1a})] and the translation
vector $\vec{R}$ discussed above.
$(M_H)_{i'i}$ in Eq.\ (\ref{i3}) is the matrix element of the $\vec{n}_H$-component of the magnetic moment $\vec{M}$
(i.e. in the direction of the magnetic field $H$), $E_j^{(0)}$ is the
energy spectrum, the ground state is described by $| i = 1 \rangle$ and the summation in Eq.~(\ref{i3}) is taken over all excited states $i'=2,3, ...\,$.
Similarly to the Langevin part, $\chi^{para}$ depends on the choice of $\vec{R}_0$, because upon the translation through $\vec{R}$
for the matrix elements $(M_H)_{i'i} = \mu_B (L_H)_{i'i}$ we obtain
\begin{eqnarray}
  \langle M'_H  \rangle = \mu_B \langle (\vec{r} + \vec{R}) \times \vec{p}\, \rangle \cdot \vec{n}_H  \nonumber \\
  = \langle M_H \rangle + \mu_B \vec{R}_{\perp} \times \langle \vec{p}_{\perp}\, \rangle \cdot \vec{n}_H ,
\label{i4}
\end{eqnarray}
(The spin contribution to $M_H$ is zero, see Sec.\ \ref{sec:VV} and Eq.\ (\ref{a4}) below.)
In Eq.~(\ref{i4}) $\langle M'_H  \rangle$ is the matrix elements in the new coordinate system,
$\vec{p} \equiv -i \vec{\nabla}$ is the operator of momentum (in atomic units) and
$\langle ... \rangle \equiv \langle i' | ... | i \rangle$.
In Eq.\ (\ref{i4}) only the perpendicular components of $\vec{R}$ and $\vec{p}$ are involved.
Thus, in diamond for $\vec{H} \parallel z$ with ${R}_{\perp} \neq 0$, Fig.\ \ref{fig1}b, the calculation of the magnetic moments $\langle M_H \rangle$
inside a displaced MT-sphere requires the matrix elements $\langle i | p_{\nu} | i' \rangle$ of momentum ($\nu=x,y$).
It is not obvious that the change in $\chi^{dia}$ e.g. in Eq.\ (\ref{i2}) is exactly compensated by the change of $\chi^{para}$
through the transformation of $\langle M'_H  \rangle$, given in Eq.\ (\ref{i4}),
especially taking into account that the Langevin contribution is solely due to the electron density (i.e. the ground state quantity)
while the Van Vleck term explicitly involves
the excited quantum states $i'$ in $\langle M'_H \rangle$.
First prove of that was sketched by Van Vleck in Ref.\ \onlinecite{VV}, where the change of $\chi^{tot}$ was considered
through laborious algebraic transformation in the energy representation for an infinitesimal increment of $\delta x$.
In Appendix A we present a compact prove of it in general form on the basis of the gauge invariance of the vector potential $\vec{A}$.
In fact, the statement of Appendix~A is a cornerstone of all present consideration, because it ensures that the contribution to $\chi$ from any unit cell $n$ of
the crystal is exactly the same as from the unit cell $n=1$ situated at the origin, Fig.\ \ref{fig1},
even though we work within a single choice of gauge for the whole crystal.

Additional magnetic contributions (``offset terms") are given by second terms on the right hand sides of Eq.~(\ref{i2}) and Eq.~(\ref{i4}).

In principle, the approach outlined here should be equally applied to the valence and core electron states.
In practice, however, it seems reasonable to use it only for the valence electrons. The main problem is numerous offset
contributions from the magnetic moment due to the nonzero matrix elements of momentum $\langle \vec{p}_{\perp}\, \rangle$
in the right hand side of Eq.\ (\ref{i4}). For example, considering the $1s$ core states of carbon, we obtain nonzero matrix elements
of the type $\langle i' |  \vec{p}_{\perp}| 1s \rangle$, where $i'$ is $np$ exited electron states ($n=$ 2, 3, \ldots).
In the case of core states, it is possible to avoid dealing with these contributions by working with the gauge origin, coinciding with the MT-sphere center
where $\vec{R}_{\perp} = 0$. (For the valence states of diamond and other solids with few atoms in the unit cell this is not possible \cite{Nik-VV}.)
With this choice of gauge, there is no contribution from the Van Vleck term, because for the $1s$ states $\langle M_H \rangle = 0$, and one has to evaluate
only the diamagnetic term. According to the gauge invariance of the sum ($\chi^{dia}_c + \chi^{para}_c$), Appendix~\ref{appA}, the same core contribution is obtained with
any other gauge origin $\vec{R}_0$.
Thus, the core contribution for carbon is calculated easily.
However, in general the core configuration can also contain $p$- and $d$- atomic states, split by an appreciable spin-orbit coupling, 
which makes the precise evaluation difficult.
For example, the silicon core configuration
is $1s^2\, 2s^2\, 2p^6$, and the magnetic moment matrix elements $\langle i' | M_H | 2p \rangle$ differ from zero for $i' \equiv np$ excited states ($n =$ 3, 4, \ldots).
For silicon therefore, the Van Vleck contribution is not zero and there appears a problem of full representation of excited states.
All these effects are beyond the scope of the present consideration and will be described elsewhere.

\section{Method: Van Vleck paramagnetism}
\label{sec:VV}

The Van Vleck paramagnetic contribution $\chi^{para}$ to the magnetic susceptibility $\chi^{tot}$, adapted for the band structure case,
can be written in the following form \cite{Nik-VV}:
\begin{eqnarray}
   \chi^{para} = 2 \sum_{\vec{k}} w(\vec{k})\, \sum_a^{N_b^o} \, \sum_p^{N_b^{uno}} \; \frac{| \langle p\, | M_H(\vec{k}) | a \rangle |^2}{E_{p}(\vec{k})-E_a(\vec{k})} .
\label{a2}
\end{eqnarray}
Here, the first summation (on $a$) is taken over all occupied bands $N_b^o$ and the second (on $p$) over unoccupied bands $N_b^{uno}$.
In general, in Eq.~(\ref{a2}) one has to integrate over all points $\vec{k}$ belonging to the first Brillouin zone. In practice the summation is taken over a set
of a representative $k-$points, Sec.\ \ref{sub:basis}, with corresponding $k-$weights $w(k)$.
For diamond the Fermi level (or rather, the chemical potential $\mu$) lies in the energy gap between $a=4$ and $p=5$ band states.
Therefore, in Eq.\ (\ref{a2}) the sum on $a$ is performed over four lowest occupied bands, i.e.
$N_b^o = 4$ and $a=1-4$, while for the sum
over unoccupied bands we have $p \geq 5$.
The upper limit for the unoccupied bands, $N_b^{uno}$, is determined by the dimension $N_b$ of the basis set used for the band structure calculation,
$N_b^{uno} = N_b - N_b^o$. (In most of our diamond calculations $N_{b}^{uno}=259$.)

The electron magnetic operator is given by $M_H \equiv M_{z'}=\mu_B(g_s s_{z'} + L_{z'} )$, where $\mu_B$ is the Bohr magneton and $g_s \approx 2$ is the electron $g-$factor.
Here the $z'$-axis correspond to the direction $\vec{n}_H$ of the external magnetic field $\vec{H}$ introduced earlier.
Note that in the absence of the spin-orbit coupling,
due to the orthogonality of the band states at each $k-$point, we have $\langle p | a \rangle = 0$ ($a \neq p$) and hence $\langle p | s_{z'} | a \rangle = 0$.
Therefore, the contribution to $\chi^{para}$ stems only from the matrix elements of the orbital momentum $L_H \equiv L_{z'}$, and
\begin{eqnarray}
   \langle p\, | M_H(\vec{k}) | a \rangle = \mu_B \langle p\, | L_H(\vec{k}) | a \rangle.
\label{a4}
\end{eqnarray}
In practice we obtain the eigenstates $| a \rangle$, $| p \rangle$ from electron band structure calculations.
In LAPW for example, one solves the eigenstate problem in the basis of augmented plane waves $\phi_j$ [see Eq.\ (\ref{m1}) below],
and, as a result, obtains the eigenstates as the coefficients of expansion in the basis set,
i.e. $\langle j | a \rangle$ and $\langle j | p \rangle$. The quantity $\langle p\, | L_H(\vec{k}) | a \rangle$
then is obtained through the matrix transformation from $\langle t\, | L_H(\vec{k}) | j \rangle$, defined in the basis of $| j \rangle$.
Therefore, the main task becomes to calculate the matrix elements of $\langle t\, | L_H(\vec{k}) | j \rangle$ in terms of basis functions $\phi_j$.
Below, we consider all different contributions to $\langle t\, | L_H(\vec{k}) | j \rangle$, comprising
$L_H^{MT}$, coming from MT-spheres, and $L_H^{IR}$ from the interstitial region.
Further, $L_H^{MT}=L_H^{MT, I} + L_H^{MT, II}$, where $L_H^{MT, II}$ is the offset term, Sec.\ \ref{sub:MT}.
Matrix elements of full magnetic moment $M_H$ in terms of $\phi_j$ are thus given by
\begin{eqnarray}
   \langle t\, | M_H(\vec{k}) | j \rangle &=& \mu_B (\langle t\, | L_H^{MT,I}(\vec{k}) | j \rangle +
   \langle t\, | L_H^{MT,II}(\vec{k}) | j \rangle  \nonumber \\
   & &+ \langle t\, | L_H^{IR}(\vec{k}) | j \rangle).
\label{a3}
\end{eqnarray}

\subsection{LAPW basis functions and computation details}
\label{sub:basis}

The LAPW method \cite{blapw,wien2k} is probably the most precise all-electron method for band structure
calculations, widely used for studies of bulk materials.

In the LAPW method space is partitioned in the region inside the nonoverlapping MT-spheres and
the interstitial region $IR$.
The basis functions $\phi_j(\vec{k},\,\vec{R})$, where $j=1,2,...,N_b$, are given by
\begin{equation}
  \phi_j(\vec{k},\,\vec{R}) = \left\{ \begin{array}{ll} V^{-1/2} \, exp(i(\vec{k} + \vec{K}_j) \vec{R}), & \vec{R} \in IR   \\
                                                    \sum_{l,m} {\cal R}_{l,m}^{j,\alpha}(r,\, E_l)\, Y_{l,m}(\hat{r}), & \vec{R} \in MT(\alpha)  \end{array} \right.
  \label{m1}
\end{equation}
where $\vec{K}_j$ refers to the reciprocal lattice vector $j$, $Y_{l,m}$ are spherical harmonics \cite{BC} and the radial part is given by
\begin{equation}
  {\cal R}_{l,m}^{j,\alpha}(r,\, E_l) = A^{j,\alpha}_{l,m}\, u_l(r, E_l) + B^{j,\alpha}_{l,m}\, \dot{u}_l(r, E_l) .
  \nonumber
\end{equation}
Here the index $\alpha$ refers to the type of atom (or
MT-sphere) in the unit cell, the radius $r$ is counted from the
center $\vec{R}_{\alpha}$ of the sphere $\alpha$ (i.e.
$\vec{r}=\vec{R}-\vec{R}_{\alpha}$), $V$ is the volume of the unit
cell. Radial functions $u_{l}(r,E_l)$ are solutions of the
Schr\"{o}dinger equation in the spherically averaged crystal
potential computed at the linearization energy $E_l$, and
$\dot{u}_l(r,E_l)$ is the derivative of $u_{l}$ with respect to
$E$ at $E_l$. The coefficients $A^{j,\alpha}_{l,m}$ and $B^{j,\alpha}_{l,m}$ are
found from the condition that the basis function $\phi_j$ is
continuous with continuous derivative at the sphere boundary,
$r=R_{MT}^{\alpha}$ ($R_{MT}^{\alpha}$ is the radius of the
MT-sphere $\alpha$):
\begin{subequations}
\begin{eqnarray}
   A^{j,\alpha}_{l,m} =  \frac{4\pi}{\sqrt{V}} i^l R_{MT}^2\, Y_{l,m}^*(\hat{k}_j)\, e^{i \vec{k}_j \vec{R}_{\alpha}}\, a^{j}_{l} , \nonumber \\
   B^{j,\alpha}_{l,m} =  \frac{4\pi}{\sqrt{V}} i^l R_{MT}^2\, Y_{l,m}^*(\hat{k}_j)\, e^{i \vec{k}_j \vec{R}_{\alpha}}\, b^{j}_{l} .  \nonumber
\end{eqnarray}
\end{subequations}
Here $\vec{k}_j = \vec{k} + \vec{K}_j$ and we have introduced the standard LAPW quantities $a^{j}_{l}$, $b^{j}_{l}$,
expressed only through the spherical Bessel functions $j_l$ and the radial solution $u_l$ (and its derivatives) at $r=R_{MT}^{\alpha}$.


For diamond there are two carbon MT-spheres in the unit cell
related through the inversion symmetry, Fig.\ \ref{fig1}.
Computation of $\chi^{dia}$, $\chi^{para}$ and $\chi^{tot}$ were performed with the core and valence electron densities
obtained as a result of the self-consistent procedure within the FLAPW band structure method \cite{blapw,wien2k},
implemented in Ref.\ \cite{Nik2}.
Diamond is crystallized in the diamond structure consisting of two simple face-centered
cubic (fcc) Bravais lattices  \cite{AM,Kit}, with two equivalent atoms in the primitive unit cell, Fig.\ \ref{fig1}.
In Fig.\ \ref{fig1} the edges of the unit cell are collinear to three basis vectors $\vec{b}_{\nu}$ ($\nu=1,2,3$) defined as in Ref.\ \cite{BC}: $\vec{b}_1=(0,1,1)\,a/2$, etc.
The technical parameters of numerical calculations for diamond are the same as those used in Ref.\ \cite{Nik-VV}.
The number of augmented plane waves was 259-283 with the wave vectors $\vec{K}_j$ satisfying the condition $R_{MT} K_j \leq 8.25$.
The maximal number of $k-$points in the irreducible part of the first Brillouin zone (BZ) was 2304, and the maximal value
of the LAPW plane-wave expansion was $l_{max} = 8$.
We have used the tetrahedron method for the linear interpolation of
energy between k points \cite{tet}. For calculation of the exchange-correlation potential and the exchange-correlation energy contribution within the DFT approach,
we have used the Perdew-Burke-Ernzerhof (PBE) variant \cite{PBE} of the generalized-gradient approximation (GGA)
and the variant of the local density approximation (LDA) with the standard ($V_{exc} \sim - \rho^{1/3}$) exchange \cite{Dir} and the PW-correlation \cite{PWcorr}.
The number of radial points inside the MT region was increased to 1000.
Finally, there were 21822 points used in the interstitial region of the unit cell.
Test calculations of the equilibrium lattice constants $a_{latt}$ and bulk moduli $B$ for diamond confirm
the results of \cite{Nik-VV}: $a_{latt}^{LDA}=3.549$~{\AA}, $B^{LDA}=474.8$~GPa (LDA) and $a_{latt}^{GGA}=3.592$~{\AA}, $B^{GGA}=440.1$~GPa (GGa),
and are in good correspondence with experimental data (3.567~{\AA} and 442~GPa).
Calculations of the magnetic susceptibility have been performed for three different lattice constants,
$a_{latt}^{LDA}$, $a_{latt}^{GGA}$, and the experimental lattice constant $a_{latt}^{exp}$.

\subsection{Magnetic Moment inside MT-spheres}
\label{sub:MT}

The expression for the matrix elements of the orbital momentum $L_H^{MT,I}(\vec{k},\alpha)$ and the corresponding magnetic moment $M_H^{MT,I} = \mu_B L_H^{MT,I}$
inside the MT-sphere $\alpha$ along $\vec{n}_H$, defined in respect to the MT-sphere center $\vec{R}_{\alpha}$,
can be obtained e.g. from equations quoted in Ref.\ \cite{Nik3}:
\begin{subequations}
\begin{eqnarray}
     \langle t\, | L_H^{MT,I} (\vec{k}, \alpha) | j \rangle  &=&  \langle t\, | \vec{L}^I(\vec{k}, \alpha) | j \rangle\, \vec{n}_H ,  \label{MT1a} \\
   \langle t\, | \vec{L}^{MT,I}(\vec{k}, \alpha) | j \rangle &=& -i \frac{4 \pi R^4_{\alpha}}{V} e^{i \vec{K}_{tj} \vec{R}_{\alpha}}\, {\cal S} \, \vec{{\cal K}} ,
\label{MT1b}
\end{eqnarray}
where $\vec{{\cal K}} = [\hat{k}_t \times \hat{k}_j]_{x,y,z}$, $\hat{k} = \vec{k}/k$, $\vec{K}_{tj} = \vec{K}_j - \vec{K}_t$ and
\begin{eqnarray}
   {\cal S} &=& \sum_l (2 l + 1) P'_l(\hat{k}_t \hat{k}_j)  \nonumber \\
  & & (a_l(\vec{k}_t)\, a_l(\vec{k}_j) + b_l(\vec{k}_t)\, b_l(\vec{k}_j) N_l) .
\label{MT1c}
\end{eqnarray}
\end{subequations}
Here $P'_l(x)$ is the derivative of the Legendre polynomial, and $N_l = \langle \dot{u}_l | \dot{u}_l \rangle$ is the radial integral of $\dot{u}_l^2(r)$
inside the MT-sphere $\alpha$ \cite{Nik3}.

However, as discussed in Sec.~\ref{sec:extra}, in reality the orbital momentum has to be evaluated in respect to another point -- the gauge origin $\vec{R}_0$, which does not
necessarily coincide with the MT-sphere center $\vec{R}_{\alpha}$.
In particular, this has to be done for any unit cell with few MT-spheres, for example, for the diamond unit cell in Fig.\ \ref{fig1}b.
In that case, as follows from Eq.\ (\ref{i4}), we should account for the extra magnetic term $\mu_B \vec{R} \times \langle \vec{p} \rangle \cdot \vec{n}_H$,
where $\langle \vec{p} \rangle$ stands for the matrix elements of the momenum inside the MT-sphere $\alpha$, i.e.
\begin{eqnarray}
 L_H^{MT} (\vec{k}, \alpha) = L_H^{MT,I} (\vec{k}, \alpha) + L_H^{MT,II} (\vec{k}, \alpha) .
\label{MT2}
\end{eqnarray}
Here the contribution $L_{H}^{MT,II} (\vec{k}, \alpha)$ is associated with the offset of the MT-sphere from the line of zero vector potential $\vec{A}$.
Matrix elements of $L_H^{MT,II} (\vec{k}, \alpha)$ are written as
\begin{eqnarray}
  \langle t\, | L_H^{MT,II} (\vec{k}, \alpha) | j \rangle  = \left[ \vec{n}_H \times \vec{R}_{\alpha} e^{i \vec{K}_{tj} \vec{R}_{\alpha}} \right]
  \langle t\, | \vec{P}(\vec{k}, \alpha) | j \rangle \, \nonumber \\
\label{MT3}
\end{eqnarray}
where $\langle \vec{P}(\vec{k}, \alpha) \rangle$ is the matrix element of momentum inside the MT-sphere $\alpha$,
\begin{eqnarray}
  & &\langle t\, | \vec{P}(\vec{k}, \alpha) | j \rangle = C_{\alpha} \sum_{l_j, \tau_j} \sum_{l_t, \tau_t} i^{l_j-l_t-1} Y_{l_t, \tau_t}(\hat{k_p}) Y^*_{l_j, \tau_j}(\hat{k_j})
   \nonumber \\
  & &\left\{ a_{l_t} a_{l_j} \, (uY_{l_t,\tau_p}| \vec{\nabla} uY_{l_j,\tau_j}) + a_{l_t} b_{l_j} \, (uY_{l_t,\tau_t}|\vec{\nabla} \dot{u}Y_{l_j,\tau_j})   \right.  \nonumber \\
  & &\left.  + b_{l_t} a_{l_j} \,  (\dot{u}Y_{l_t,\tau_t}|\vec{\nabla} uY_{l_j,\tau_j})
          + b_{l_t} b_{l_j} \, (\dot{u}Y_{l_t,\tau_t}|\vec{\nabla} \dot{u}Y_{l_j}) \right\}.
\label{MT4}
\end{eqnarray}
Here $C_{\alpha} = (4 \pi)^2 R_{\alpha}^4 / V$, index $\tau$ refers to $m$ or $(m,c)$, $(m,s)$ (in case of cos- and sin-like real spherical harmonics \cite{BC})
and the integrals are given be
\begin{eqnarray}
 (uY_{l_1,\tau_1}|\vec{\nabla} uY_{l_2,\tau_2}) = \int_0^{R_{MT}^{\alpha}} \int_{\Omega} u_{l_1}(r)\, Y^*_{l_1,\tau_1}(\Omega) \nonumber \\
 \vec{\nabla} \left\{ u_{l_2}(r)\, Y_{l_1,\tau_1}(\Omega) \right\} r^2 dr \, d\Omega .
   \label{MT5}
\end{eqnarray}
Although in Eq.\ (\ref{MT4}) in principle we deal with a double sum, Eq.\ (\ref{MT5}) acts as a selection rule and greatly reduces the number of
non-zero terms in Eq.\ (\ref{MT4}).
In fact, the non-zero terms are only those for which
\begin{eqnarray}
 l_t = l_j \pm 1, \nonumber
\end{eqnarray}
which is imposed by the parity consideration.

Explicit expressions for the integrals $(uY_{l_1,\tau_1}|\vec{\nabla} uY_{l_2,\tau_2})$ can be worked out with the help of
equations in Sec.\ 5.8.3 of Ref.\ \cite{Varsh} (see also Appendix~A of \cite{Laskow12}), noting that $\partial_x = (-\nabla_{+1} + \nabla_{-1})/\sqrt{2}$,
$\partial_y = i(\nabla_{+1} + \nabla_{-1})/\sqrt{2}$ and $\partial_z = \nabla_{0}$.
Since the operator of momentum is hermitian,
Eq.\ (\ref{MT4}) should be symmetrized in respect to $\vec{\nabla}$ operating on the functions standing on the right and left sides of Eq.\ (\ref{MT4}).
(The same result is obtained by taking into full account the integration on the surface region of MT-sphere.)

It is worth considering these equations for a system having the inversion symmetry, for example, for the diamond, Fig.\ \ref{fig1}.
In that case there are two equivalent MT-spheres, and the expression $\exp(i \vec{K}_{tj} \vec{R}_{\alpha})$ in Eq.\ (\ref{MT1b}) and Eq.\ (\ref{MT3})
should be replaced by the sum over two MT-spheres (i.e. $\alpha=1,2$) with $\vec{R}_{\alpha=1,2} = \pm \vec{R}_1$, where $\vec{R}_1=(1,1,1)\,a_{latt}/8$.
In Eq.\ (\ref{MT1b}) the summation gives the standard LAPW structure factor $\sum_{\alpha=1}^2 \exp(i \vec{K}_{tj} \vec{R}_{\alpha})= 2 \cos(\vec{K}_{tj} \vec{R}_1)$.
Thus, the matrix elements of $L_H^I$, Eqs.\ (\ref{MT1a})--(\ref{MT1c}), and consequently $M_H^I$, are imaginary.
The same consideration, applied to the term in square brackets of Eq.\ (\ref{MT3}), yields $[\vec{n}_H \times \vec{\cal R}]$,
where
\begin{eqnarray}
 \vec{\cal R} = \sum_{\alpha=1}^2 \vec{R}_{\alpha} e^{i \vec{K}_{tj}} = 2\, i\, \vec{R}_1 \sin(\vec{K}_{tj} \vec{R}_1) .
   \label{MT7}
\end{eqnarray}
Since in terms of real spherical harmonics $\langle \vec{P}(\vec{k}, \alpha) \rangle$, Eq.\ (\ref{MT4}), is real,
the matrix elements of $L_H^{MT,II}$, Eq.\ (\ref{MT3}), and $M_H^{MT,II}$ are imaginary.
Consequently, the matrix elements of $L_H^{MT}$, Eq.\ (\ref{MT2}), and the full magnetic moment $M_H^{MT}= \mu_B L_H^{MT}$ inside the MT-spheres are also imaginary.

\subsection{Van Vleck contribution from the interstitial region}
\label{sub:VVI}

Here we calculate the matrix elements of $L_H^{IR}$ and $M_H^{IR}$ for the interstitial region (IR).
Starting with the plane wave expression for the LAPW basis functions, Eq.\ (\ref{m1}), and using $\vec{L}=-i [\vec{r} \times \vec{\nabla}]$ we obtain
\begin{eqnarray}
 \langle t\, | L_H^{IR} (\vec{k}) | j \rangle = \frac{1}{V} \int_{IR} e^{i \vec{K}_{tj} \vec{r}}\, [\vec{r} \times (\vec{k} + \vec{K}_j) ]\, d\vec{r} ,\;
   \label{ir1}
\end{eqnarray}
where the integration is taken over the interstitial region.
Taking into account the hermicity of $\vec{L}$ (or, equivalently, treating accurately the MT-sphere surface integration),
and adapting the situation to the unit cell inversion symmetry, Eq.\ (\ref{ir1}) can be rewritten as
\begin{subequations}
\begin{eqnarray}
 \langle t\, | L_H^{IR} (\vec{k}) | j \rangle = \left[ \vec{n}_H \times \vec{\cal I}_{\,tj} \right]
 \left( \vec{k} + \frac{1}{2}(\vec{K}_t + \vec{K}_j) \right) , \quad
  \label{ir2a}
\end{eqnarray}
where
\begin{eqnarray}
 \vec{\cal I}_{\,tj} = \frac{1}{V} \int_{IR} e^{i \vec{K}_{tj}\, \vec{r}}\, \vec{r} \, d\vec{r} .
  \label{ir2b}
\end{eqnarray}
\end{subequations}
The integral $\vec{\cal I}_{\,tj}$, Eq.\ (\ref{ir2b}), comprises three components ${\cal I}_{\,tj,\, x}$, ${\cal I}_{\,tj,\, y}$ and ${\cal I}_{\,tj,\, z}$,
which can be found in analytical form from
\begin{subequations}
\begin{eqnarray}
  {\cal I}_{\,tj,\,x} = -i \, \left( \frac{\partial}{\partial K_{tj,\, x}}\; O_{\,tj}^{IR} \right)  ,
  \label{ir3a}
\end{eqnarray}
and analogous relations for the $y$- and $z$-components. Here
\begin{eqnarray}
  O_{\,tj}^{IR} = \frac{1}{V} \int_{IR} e^{i \vec{K}_{tj}\, \vec{r}}\, d\vec{r}
  \label{ir3b}
\end{eqnarray}
\end{subequations}
is the well known LAPW overlap integral in the interstitial region.
It is worth mentioning that Eq.\ (\ref{ir3a}) and Eq.\ (\ref{ir3b}) should be treated with care -- for example,
cases with $O_{\,tj}^{IR}=0$ can result in $\partial O_{\,tj}^{IR} / \partial K_{tj,\, x} \neq 0$.
More details on analytic expressions for $\vec{\cal I}_{\,tj}$ are given in Appendix~B.

Finally, we note that for diamond and other cases with inversion symmetry, the overlap integrals $O_{\,tj}^{IR}$ are real
while the components of $\vec{\cal I}_{\,tj}$ are imaginary, leading to imaginary matrix elements for $L_H^{IR}$ and $M_H^{IR}$, Eq.\ (\ref{ir1}).

\section{Method: Langevin diamagnetism}
\label{sec:LL}

In contrast to the Van Vleck paramagnetism operating with wave functions of states and their energies,
the evaluation of the Langevin (Larmor) diamagnetism is based only on the knowledge of
the electron densities ($\rho_{val}(\vec{r})$ and $\rho_{c}(\vec{r})$ of the valence and core electrons, respectively) and the
direction $\vec{n}_H$ of the applied magnetic field $H$.
As discussed in Sec.\ \ref{sec:extra}, in calculating the diamagnetic response from the interstitial region we should use
direct equations (\ref{i1}), (\ref{i1b}) with perpendicular component of radius-vector $r_{\perp}$, counted from $\vec{R}_0 = 0$.

The same holds for the calculation of the MT-sphere contribution $\langle {r}_{\perp}^2 \rangle^{MT}_{val}$ from the valence electrons.
However, from the technical point of view it is more convenient to compute first the quantity $\chi^{dia}_{MT,0}=-C_L \langle (\vec{r}-\vec{R}_c)_{\perp}^2 \rangle^{MT}_{val}$
with $r_{\perp}$, counted from the MT-sphere center $\vec{R}_c$. In that case, as follows from Eq.\ (\ref{i2}), in addition to $\chi^{dia}_{MT,0}$
the diamagnetic response $\chi^{dia}_{MT}$
acquires also the extra (offset) contribution $\chi^{dia}_{MT,off}=-C_L R_{c,\perp}^2 |Q^{MT}_{val}|$.
(We recall that $R_{c,\perp}$ is the perpendicular component of $\vec{R}_c$,
and $Q^{MT}_{val}$ is the total charge of the valence electrons inside the sphere.)

In the following we apply our method to diamond. Taking into account that there are two MT-spheres in the unit cell, for the MT valence contribution
we obtain
\begin{subequations}
\begin{eqnarray}
   \chi^{dia}_{MT,v} = 2\, \chi^{dia}_{MT,0,v} + \chi^{dia}_{MT,off, v} ,
  \label{l1}
\end{eqnarray}
where the offset term is given by
\begin{eqnarray}
   \chi^{dia}_{MT,off, v} = - 2\, C_L R_{c,\perp}^2 |Q^{MT}_{val}| .
  \label{l1b}
\end{eqnarray}
\end{subequations}
Here $\chi^{dia}_{MT,0,\, v}$ is the diamagnetic contribution of the valence electrons from a single MT-sphere with $r_{\perp}$ counted from the sphere center,
which can be rewritten in the familiar form,
\begin{eqnarray}
   \chi^{dia}_{MT,0,\, v} = -\frac{2}{3}C_L \langle r^2 \rangle_{val} ,
  \label{l2}
\end{eqnarray}
where $\langle r^2 \rangle_{val}$ is the value of $r^2$ for valence electrons averaged inside the sphere with the
spherically symmetric component $\rho_{val}^{L=0}(r)$ of the valence density. For diamond the high multipole moments with $L > 2$
do not contribute to $\chi^{dia}_{MT,0,\, v}$. Indeed, $r_{\perp}^2 = c_0 Y_{L=0,0} + c_{\tau} Y_{L=2,\tau}$ and from symmetry it follows that
the nonzero integral over $r_{\perp}^2$ can be only with density components with $L = 0,\, 2$, whereas in diamond the first non-zero
density multipole is an octupole with $L=3$ \cite{BC}.

In principle, Eq.\ (\ref{l1}) should also be employed for the diamagnetic response $\chi_{MT,\,c}$ from the core electrons.
However, as discussed in the end of Sec.~\ref{sec:extra}, the core contribution can be calculated more easily
chosing the gauge origin, coinciding with the MT-sphere center.
For core diamagnetic part we then obtain $\chi^{dia}_c = \chi^{dia}_{MT,0,\, c}$,
with the paramagnetic counterpart $\chi^{para}_c = 0$, because the orbital magnetic moment for the $1s$ core states is zero.
The total magnetic response from the core electrons is
\begin{eqnarray}
   \chi_{MT,\,c} = 2\, \chi^{dia}_{MT,0,\, c} ,
  \label{l3}
\end{eqnarray}
where $\chi^{dia}_{MT,0,\, c}$ is given by Eq.\ (\ref{l2}) with $\langle r^2 \rangle_{val}$ substituted by $\langle r^2 \rangle_{c}$.

\section{Diamagnetism and anisotropy of $\chi$ in diamond}
\label{sec:diamond}

Here we consider the magnetic susceptibility of pristine diamond applying the formulas from Sec.\ \ref{sec:VV}. and Sec.\ \ref{sec:LL} to
the calculation of $\chi^{dia}$, $\chi^{para}$ and $\chi^{tot}$.

One of the most important consequences of the induced magnetization, caused by the external magnetic field $H$ is
the emergence of a new crystalline symmetry, which takes into account the direction $\vec{n}_H$ of $\vec{H}$.
For example, if $\vec{n}_H$ is collinear to the [111] axis, the initial cubic symmetry is reduced to the three-fold rotation symmetry $C_{3v}$
about [111]. If $\vec{n}_H$ is parallel to the [001] axis, the new symmetry is the four-fold rotation $C_{4v}$ about [001], etc.
For the general direction of $\vec{n}_H$ ($n_{H,x} < n_{H,y} < n_{H,z}$) there is no crystal point symmetry and the
only remaining symmetry operation is inversion.
In our calculations we have observed these effects explicitly.

\subsection{Langevin diamagnetic contributions}
\label{sec:diaL}

First, we consider the Langevin diamagnetic response $\chi^{dia}$, which has two main contributions -- from the valence and core electrons,
$\chi^{dia}$(tot.)$=\chi^L$(val.)$+\chi^L$(core). The core contribution is given by the MT value $\chi^L$(core)$=\chi_{MT,\,c}$, Eq.\ (\ref{l3}), our
LDA calculation yields $-0.468 \cdot 10^{-7}$ \cite{Nik-VV} for all $\vec{n}_H$.
In turn, the valence contribution consists of two parts, $\chi^L$(val.)$=\chi^L$(IR)$+\chi^L$(MT),
from the interstitial region (IR) and MT-spheres (MT).
The latter, Eq.\ (\ref{l1}), includes also the offset term $\chi^L$(offset), Eq.\ (\ref{l1b}).
In the interstitial region the value of $r^2_{\perp}$ is directly determined by $\vec{n}_H$,
which explicitly changes the integral $\langle r^2_{\perp} \rangle$, Eq.\ (\ref{i1b}).
The dependence of some contributions on $\vec{n}_H$ is demonstrated in Table~\ref{tab1}.
(The directions $[-a_1,a_1,b_1]$ and $[a_2,b_2,0]$ in Table \ref{tab1}, belong to the 74 point Lebedev grid \cite{Leb1,Leb2},
used for a precise surface integration; $a_1=0.480384$, $b_1=0.733799$, and $a_2=0.320773$, $b_2=0.947156$).
%
\begin{table}
\caption{
Contributions $\chi^{L}$ to the Langevin diamagnetic response $\chi^{dia}$ (tot.) for various directions of $\vec{H}$
from the valence electrons (val.), interstitial region (IR), and the offset term, Eq.\ (\ref{l1b}),
LDA calculations with $a_{latt}^{exp}$; all $\chi$ are volume values,
in units $10^{-7}$, see text for details.
\label{tab1} }
\begin{ruledtabular}
\begin{tabular}{c  c  c  c  c}
 $\vec{H}\; \parallel$  &  $\chi^{L}$, val.  & $\chi^{L}$, IR & $\chi^{L}$, offset & $\chi^{dia}$, tot.  \\
\tableline
  $[1,1,1]$  & -18.146 & -11.452 & 0 & -18.614 \\
  $[1,1,-1]$ & -46.207 & -39.513 & -16.356 & -46.675 \\
  $[0,0,1]$  & -39.192 & -32.498 & -12.267 & -39.659 \\
  $[1,-1,0]$ & -49.714 & -43.020 & -18.400 & -50.182 \\
  $[-a_1,a_1,b_1]$ & -44.048 & -37.354 & -15.097 & -44.516 \\
  $[a_2,b_2,0]$  & -32.798 & -26.104 & -8.540 & -33.265 \\
\end{tabular}
\end{ruledtabular}
\end{table}
The $\vec{n}_H$-independent MT valence contribution, determined by Eq.\ (\ref{l2}), is $-6.694 \cdot 10^{-7}$ (LDA).
Thus, the dependence of $\chi^{dia}_{MT,v}$ on $\vec{n}_H$ is solely due to the offset term $\chi^{dia}_{MT,off,v}$, Eq.\ (\ref{l1b}),
which explicitly depends on $R_{c,\perp}$.

\begin{figure}
\resizebox{0.48\textwidth}{!} {
\includegraphics{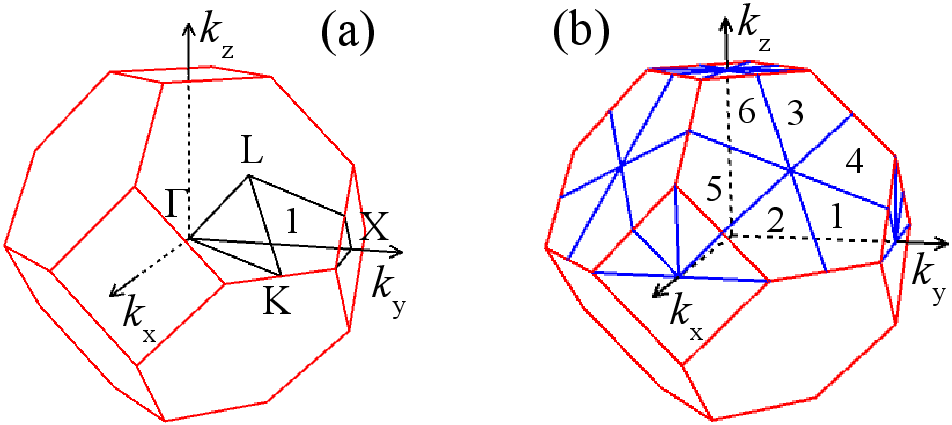}
}

\caption{
The Brillouin zone of the diamond structure in reciprocal ($k$) space.
(a) Irreducible part of BZ (1/48) without external magnetic field;
(b) Irreducible part of BZ (upper half, 1/2) in the presence of magnetic field,
divided into 24 subsections, each of which is equivalent to the part 1 shown in (a).
} \label{fig2}
\end{figure}

It is worth mentioning that after fixing $\vec{n}_H$ the calculation of all Langevin contributions requires only the valence ($\rho_{val}$) and
core ($\rho_{c}$) electron density,
taken from band structure calculations. Both $\rho_{val}$ and $\rho_c$ are obtained using the standard irreducible part of the Brillouin zone.
For diamond the irreducible part, shown in Fig.\ \ref{fig2}a, is only 1/48 part of the whole Brillouin zone.
Thus, the diamagnetic Langevin contribution $\chi^{dia}$ can be obtained using the band structure calculations with the standard (1/48) irreducible
part of the Brillouin zone. The situation is very different for the Van Vleck paramagnetic response $\chi^{para}$ considered below.
%
%

\subsection{Enhanced irreducible part of the Brillouin zone in calculations of Van Vleck contributions}
\label{sec:enhance}

To deal with various $k-$point contributions to $\chi^{para}$, Eq.\ (\ref{a2}), we will partition the total $k-$sum in Eq.\ (\ref{a2}) in
subsums, each of which is limited to a certain part of the Brillouin zone, shown schematically in Fig.\ \ref{fig2}b.
We will classify these parts as follows. The first six parts, having the same volume ($V_k/48$) in $k-$space, are shown in Fig.\ \ref{fig2}b explicitly.
The other 18 parts are obtained from these six ones by rotations by angles $\pi/2$, $\pi$, $3\pi/2$ about the [001] four-fold rotation axis.
Correspondingly, we will denote them as ${\cal N}_r$, where $n=1,2,...,6$ refers to the parts shown in Fig.\ \ref{fig2}b, and $r=0,1,2,3$ refers to
the subsequent rotation about the [0,0,1]-axis by the angle $\varphi=0$, $\pi/2$, $\pi$, $3\pi/2$.
These 24 subregions of ${\cal N}_r$ make up the upper half of the entire Brillouin zone, and given that the inversion symmetry is preserved in an applied magnetic field,
this is sufficient for a complete description.
Our study shows that indeed for the calculations of the Van Vleck contribution along some directions
of $H$, summations over all 24 subparts ${\cal N}_r$ are required, details are given in Appendix~\ref{appC}.
Thus, the irreducible part of the Brillouin zone for diamond in an external magnetic field includes the entire upper half of the zone, shown in Fig.\ \ref{fig2}b.

\subsection{Analyses of $\chi^{tot}$ for various directions of \\magnetic field}
\label{sub:dir}

In this subsection we analyze the resulting total magnetic susceptibility
$\chi^{tot}(\vec{n}_H)=\chi^{dia}(\vec{n}_H) + \chi^{para}(\vec{n}_H)$ for various directions $\vec{n}_H$ of magnetic field $H$.
The results for some selected values of $\vec{n}_H$ are reproduced in Table \ref{tab4}.
A comparative study of partial contributions to $\chi^{para}$ from the interstitial region and MT-sphere region is given in Appendix~\ref{appD}.

We first discuss the case of $\vec{H}$ parallel to the $z-$axis, $\vec{n}_H = (0,0,1 )$.
The same contributions $\chi^{tot}$, $\chi^{dia}$ and $\chi^{para}$ occur for $\vec{H}$ parallel
to the $x-$axis and $y-$axis. This is a consequence of the symmetry of the chosen unit cell, Fig.\ \ref{fig1},  and the Brillouin zone, Fig.\ \ref{fig2}.
Indeed, the subsequent counterclockwise rotations by $2 \pi /3$ about the [1,1,1]-axis transform the unit cell and Brillouin zone
into themselves, whereas the $z-$axis transforms to the $x-$axis and then to the $y-$axis.
%
\begin{table}
\caption{
Full diamagnetic (Langevin) response $\chi^{dia}$, and full paramagnetic (Van Vleck) response $\chi^{dia}$, and their
sum $\chi^{tot}$ for selected directions $\vec{n}_H$ of $\vec{H}$. The subscript $av$ of $\vec{n}_H$ stands for $\chi^{dia}$,
$\chi^{para}$, $\chi^{tot}$ averaged over all equivalent directions $\vec{n}'_H$, see text for details. LDA calculations with
$a_{latt}^{LDA}$, all $\chi$ are volume values, in units $10^{-7}$.
\label{tab4} }
\begin{ruledtabular}
\begin{tabular}{l   r  r  r   r  r  r}
       &   \multicolumn{3}{c}{$\chi$, LDA} & \multicolumn{3}{c}{$\chi$, GGA}   \\
 $\vec{H} \parallel$ & dia  & para  & tot  & dia  & para  & tot  \\
\tableline
  $[0,0,1]$         & -39.662 & 23.136 & -16.526 & -39.598 & 23.054 & -16.544 \\
  $[0,0,1]_{av}$    & -39.662 & 23.136 & -16.526 & -39.598 & 23.054 & -16.544 \\
\tableline
  $[1,1,1]$         & -18.616 & 10.712 &  -7.904 & -18.592 & 10.662 & -7.930 \\
  $[1,1,-1]$        & -46.678 & 27.276 & -19.402 & -46.600 & 27.185 & -19.415 \\
  $[1,1,1]_{av}$    & -39.662 & 23.136 & -16.526 & -39.598 & 23.054 & -16.544 \\
\tableline
  $[1,1,0]$         & -29.139 & 16.924 & -12.215 & -29.095 & 16.858 & -12.237 \\
  $[1,-1,0]$        & -50.186 & 29.348 & -20.838 & -50.101 & 29.250 & -20.851 \\
  $[1,1,0]_{av}$    & -39.662 & 23.136 & -16.526 & -39.598 & 23.054 & -16.544 \\
\tableline
  $[a_1,a_1,b_1]$   & -19.968 & 11.510 &  -8.458 & -19.941 & 11.458 & -8.483 \\
  $[-a_1,a_1,b_1]$  & -44.519 & 26.003 & -18.516 & -44.445 & 25.914 & -18.531 \\
  $[-a_1,-a_1,b_1]$ & -49.643 & 29.028 & -20.615 & -49.560 & 28.931 & -20.629 \\
$[a_1,a_1,b_1]_{av}$& -39.662 & 23.136 & -16.526 & -39.598 & 23.054 & -16.544 \\
\tableline
  $[a_2,b_2,0]$     & -33.268 & 19.361 & -13.907 & -33.216 & 19.289 & -13.927 \\
  $[-a_2,b_2,0]$    & -46.056 & 26.910 & -19.146 & -45.980 & 26.819 & -19.161 \\
 $[a_2,b_2,0]_{av}$ & -39.662 & 23.136 & -16.526 & -39.598 & 23.054 & -16.544 \\

\end{tabular}
\end{ruledtabular}
\end{table}

However, if we consider $H$ applied along main cube diagonals (i.e. [1,1,1], $[1,1,-1]$, $[-1,1,1]$ and $[1,-1,1])$, the situation is different.
In that case, $\chi^{tot}(\vec{n}_H)$, $\chi^{dia}(\vec{n}_H)$ and $\chi^{para}(\vec{n}_H)$, found for these directions, are not equal.
In Table \ref{tab4} we quote these values for [1,1,1] and $[1,1,-1]$.
Note, that the rotation about the [1,1,1]-axis transforms the $[1,1,-1]$-axis first to $[-1,1,1]$,
and then to $[1,-1,1]$.
Thus, the values of $\chi^{tot}(\vec{n}_H)$, $\chi^{dia}(\vec{n}_H)$ and $\chi^{para}(\vec{n}_H)$ should be the same for $[1,1,-1]$, $[-1,1,1]$ and $[1,-1,1]$,
which was confirmed by our direct calculations.
Since in real diamond crystal these directions are equivalent, we consider the average value of $\chi^{tot}_{av}$, $\chi^{dia}_{av}$ and $\chi^{para}_{av}$, calculated according to the occurrence of individual values, i.e.
\begin{eqnarray}
   \chi[1,1,1]_{av} = ( \chi[1,1,1] + 3\, \chi[1,1,-1] ) / 4 .
  \label{r1}
\end{eqnarray}
This average value is given in Table \ref{tab4} as $[1,1,1]_{av}$.
Note, that it exactly corresponds to $[0,0,1]_{av}$ (i.e when $H$ is parallel to [0,0,1], [1,0,0] or [0,1,0]).
In fact, as follows from Table \ref{tab4}, the average values $\chi^{tot}_{av}$, $\chi^{dia}_{av}$ and $\chi^{para}_{av}$ are the same
for any direction $\vec{n}_H$ of the magnetic field: they represent the actual magnetic susceptibility and its full diamagnetic and paramagnetic contribution.

In Table \ref{tab4} this conclusion has been checked for six equivalent directions $[1,1,0]$, $[1,-1,0]$, etc.\! with two non-equivalent $\chi$-values,
for 12 equivalent directions $[a_1,a_1,b_1]$, $[a_1,b_1,a_1]$ etc.\! with three non-equivalent $\chi$-values, and
for 12 equivalent directions $[a_2,b_2,0]$, $[a_2,0,b_2]$ etc.\! with two non-equivalent $\chi$-values.
Here two groups of $\vec{n}_H$-directions -- $[a_1,a_1,b_1]$ and $[a_2,b_2,0]$ -- belong to the 74 point Lebedev grid \cite{Leb1,Leb2},
used for accurate surface integration, where $a_1 \approx 0.480384$, $b_1=\sqrt{1 - 2 a_1^2}$,
and $a_2 \approx 0.320773$, $b_2 = \sqrt{1 - a_2^2}$.

The procedure of averaging used in this subsection, can be viewed as an analogue of the symmetrization of the electron density
obtained at a $\vec{k}$-point in accordance with its full crystal symmetry.

\subsection{Invariance of $\chi^{tot}$, $\chi^{dia}$ and $\chi^{para}$ in respect to the direction of magnetic field}
\label{sub:inv}

If the magnetic susceptibility $\chi^{tot}(\vec{n}_H)$ is anisotropic in respect to the direction $\vec{n}_H$ of magnetic field,
defined by the polar angles $\vec{n}_H \equiv (\theta, \varphi)$, it can be expanded in multipolar series,
\begin{eqnarray}
   \chi^{tot}(\theta, \varphi) = \chi_0 + \chi_4 K_{L=4}(\theta, \varphi) + \chi_6 K_{L=6}(\theta, \varphi), \quad
  \label{r2}
\end{eqnarray}
where $\chi_0$ is the isotropic part, while the anisotropy is described by the cubic harmonics $K_{L=4}$, $K_{L=6}$, which are
symmetry adapted combinations of spherical harmonics with $L=4$ and 6, which are invariant under symmetry operations of
the tetrahedral ($T_d$) point symmetry. The explicit form of $K_{L=4}$, $K_{L=6}$ can be found e.g. in \cite{BC}),
and in Eq.\ (\ref{r2}) we have ignored the high multipole terms with $L \geq 8$.
The coefficients $\chi_4$ and $\chi_6$ are found from
\begin{eqnarray}
  \chi_L = \int_{\Omega} d \Omega \, \chi^{tot}(\theta, \varphi) \, K_{L}(\theta, \varphi)  \nonumber \\
     \rightarrow \sum_i w_i \chi^{tot}(\theta_i, \varphi_i) \, K_{L}(\theta_i, \varphi_i) ,
  \label{r3}
\end{eqnarray}
where the integration on $\Omega=(\theta, \varphi)$ is replaced by summation on points $(\theta_i, \varphi_i)$ with the weight $w_i$.
To perform the summation  in Eq.\ (\ref{r3}), we have used the 74 point Lebedev surface grid \cite{Leb1,Leb2}, which is enough to extract
the coefficients $\chi_{L=4}$ and $\chi_{L=6}$.
Using results listed in Table \ref{tab4},
we find that $\chi_4 = \chi_6 = 0$, and the only nonzero term in Eq.\ (\ref{r2}) is $\chi_0$.
The same result holds for $\chi^{dia}$ and $\chi^{para}$.
Therefore, all magnetic susceptibilities ($\chi^{tot}$, $\chi^{dia}$ and $\chi^{para}$) are independent of the direction $\vec{n}_H$
of the magnetic field $H$. The actual values of $\chi^{tot}$, $\chi^{dia}$ and $\chi^{para}$, obtained within the LDA and GGA variants of DFT,
are collected in Table \ref{tab5}.
%
\begin{table}
\caption{
DFT results for $\chi^{dia}$, $\chi^{para}$, $\chi^{tot}$ (volume values, in $10^{-7}$) for different lattice constants $a_{latt}$:
(1) equilibrium LDA lattice constant $a_{latt}^{LDA}$; (2) equilibrium GGA lattice constant $a_{latt}^{GGA}$; (3) experimental lattice constant $a_{latt}^{exp}$.
(*) and (**): experimental volume susceptibilities $\chi^{exp}$, obtained from
(*) density value $\chi_{\rho}^{exp}=-4.5 \times 10^{-7}$~cm$^3$/g \cite{Her};
(**) molar value $\chi_{m}^{exp}=-5.9 \times 10^{-6}$~cm$^3$/mole \cite{Hud}. ($a$) equilibrium $a_{latt}$ in Ref.~\cite{ML96}.
\label{tab5} }
\begin{ruledtabular}
\begin{tabular}{c  c  c  c  c}
 type DFT  &  $a_{latt}$ &  $\chi_{dia}$ &  $\chi_{para}$ & $\chi_{tot}$  \\
\tableline
  LDA      & 3.549$^{(1)}$ {\AA} & -39.940 & 23.219 & -16.721 \\
  GGA      & 3.592$^{(2)}$ {\AA} & -39.216 & 22.942 & -16.274 \\
  LDA      & 3.567$^{(3)}$ {\AA} & -39.662 & 23.136 & -16.526 \\
  GGA      & 3.567$^{(3)}$ {\AA} & -39.598 & 23.054 & -16.544 \\
\tableline
  LDA \cite{ML96}     & 3.52$^{\,a}$ {\AA} &         &        &  -16.35 \\
  LDA \cite{ML96}     & 3.57$^{(3)}$ {\AA} &         &        &  -16.44 \\
  exp.* \cite{Her}    &      &         &        & -15.82 \\
  exp.** \cite{Hud}   &      &         &        & -17.27 \\
\end{tabular}
\end{ruledtabular}
\end{table}

In fact, the independence from $\vec{n}_H$ is already present in Table \ref{tab4}, manifesting itself in equal averaged values of $\chi_{av}$.
The integration, Eq.\ (\ref{r3}), ascertains that these averaged values are equal to the actual ones, Table \ref{tab5} (the LDA variant at $a_{latt}^{exp}$).

\begin{figure}
\resizebox{0.30\textwidth}{!} {
\includegraphics{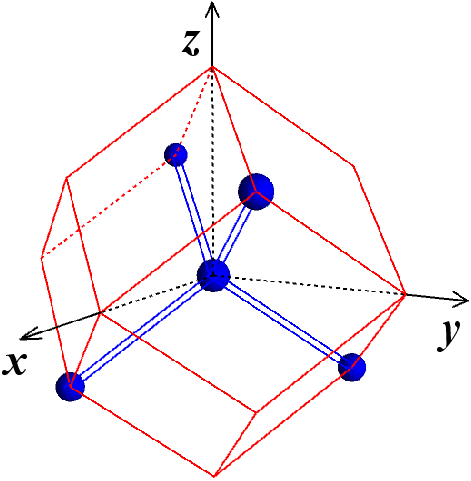}
}

\caption{
The fully symmetrical (tetrahedral) diamond unit cell as a Wigner-Seitz primitive unit cell. One carbon atom is at the center (with weight $w = 1$),
the four remaining atoms lie at vertices ($w=1/4$);
together they represent the two carbon atoms in the unit cell (compare with Fig.\ \ref{fig1}).
The double blue lines represent C-C carbon bonds.
} \label{fig3}
\end{figure}

The question may arise: why $\chi^{tot}$, $\chi^{dia}$ and $\chi^{para}$ are different for some individual directions of $\vec{H}$, for example, for $[1,1,1]$ and $[1,1,-1]$,
Table \ref{tab4}, although these crystal axes should be equivalent? We think that this is directly connected with the choice of the primitive unit cell.
If we consider the primitive unit cell, chosen by us, Fig.~\ref{fig1}, we find that its shape is not invariant under all symmetry operations
of the diamond crystal lattice.
This does not matter for electron band structure calculations because we impose the symmetry restrictions on the electron density by expanding it only
in symmetry adapted functions inside the MT-spheres and in stars of $\vec{K}_j$ in the interstitial region \cite{blapw}.
In the magnetic calculation, the symmetry restrictions are equivalent to the introduced procedure of averaging,
e.g. in Eq.\ (\ref{r1}).
The equivalence of the individual directions $[1,1,1]$ and $[1,-1,1]$ and all others, can be explicitly restored if one uses the primitive unit cell,
reproduced in Fig.\ \ref{fig3}. This is the Wigner-Seitz unit cell having the full tetrahedral site symmetry.
Although this choice of unit cell looks very attractive, it removes the inversion and requires working with partial occupancy of four carbon atoms, which is inconvenient in practical calculations.
%
%

\section{Conclusions}
\label{sec:con}

We have presented a general method for the calculation of all contributions to the Van Vleck paramagnetism and the Langevin diamagnetism and
applied it to the calculation of the magnetic susceptibility $\chi^{tot}$ of diamond, which is a very important technological material.
In contrast to other approaches \cite{ML96,MPL,Gregor99,Pickard01,Laskow12,Laskow14}, based on the initial linear response formulation of Mauri et al. \cite{ML96,MPL},
the method (1) uses a single choice of the gauge for valence electrons throughout the whole crystal, and (2) works in the static external magnetic field $H$,
i.e. no modulation with a small wave vector $q$ of $H$ is assumed, $q = 0$.
The consideration is based on the statement (Appendix~A) that the sum of the Van Vleck term, $\chi^{para}$, and the Langevin term, $\chi^{dia}$,
constituting the total susceptibility $\chi^{tot}= \chi^{para} + \chi^{dia}$, is independent of the choice of the magnetic gauge origin,
although individually $\chi^{para}$ and $\chi^{dia}$ can demonstrate such dependence.

The method is adapted for the LAPW basis functions, which are plane waves in the interstitial region, and atomic-like wave functions in the MT-sphere region, Eq.\ (\ref{m1}).
Since we use a single choice of magnetic gauge for the whole crystal, we obtain unfamiliar contributions which we call offset terms.
For example, such a term is given by the magnetic moment $M_H^{MT,II}$, Eq.\ (\ref{MT3}),
appearing when the MT-sphere center is displaced from the gauge origin.
Likewise, the offset term is present in the Langevin calculation of $\chi^{dia}$, Eq.\ (\ref{l1b}).
Finally, we mention a contribution from the interstitial region, Sec.\ \ref{sub:VVI}.
All magnetic contributions are obtained in analytical form and then implemented in our LAPW code \cite{Nik2}.
The method can be applied to any solid including dielectrics and metals. In metals it can be used for
obtaining the Langevin contribution $\chi^{dia}$ and the Van Vleck contribution $\chi^{para}$ in addition to
the well known Pauli paramagnetism and the Landau diamagnetism.

Applying this method to diamond, we have performed FLAPW calculations of $\chi^{para}$, $\chi^{dia}$ and $\chi^{tot}$,
analyzing various partial contributions for different directions of the applied magnetic field $H$.
We have found that
the application of the magnetic field changes the crystal symmetry, which manifests itself in the enhanced irreducible part of the Brillouin zone,
required for such calculations, Sec.\ \ref{sec:enhance} and Fig.\ \ref{fig2}.

In study of diamond our goal was to extract the anisotropic part, described by the cubic harmonics $K_{L=4}(\vec{n}_H)$, $K_{L=6}(\vec{n}_H)$
in $\chi^{tot}$, Eq.\ (\ref{r2}).
For that purpose we have performed integration in Eq.\ (\ref{r3}), using the 74 point Lebedev surface grid \cite{Leb1,Leb2},
and found that the anisotropic coefficients $\chi_4$, $\chi_6$ in Eq.\ (\ref{r2}), are reduced to zero, i.e. $\chi_4 = \chi_6 = 0$.
Note, however, that the calculated individual susceptibilities $\chi^{para}$, $\chi^{dia}$ and $\chi^{tot}$ depend on the direction of $H$.
We consider that as an artefact of the calculations, related to a relatively low symmetry of the chosen primitive unit cell, Fig.\ \ref{fig1}.
In Sec.\ \ref{sub:inv} we argue that the individual susceptibilities $\chi^{para}$, $\chi^{dia}$ will be independent of $\vec{n}_H$
for the fully symmetric Wigner-Seitz primitive unit cell of diamond, Fig.\ \ref{fig3}.

Thus, our calculations result in magnetic susceptibilities $\chi^{para}$, $\chi^{dia}$ and $\chi^{tot}$, quoted in Table \ref{tab5},
which are independent of the direction $\vec{n}_H$ of the applied magnetic field $H$.
The calculated values $\chi^{tot}$ in other units lie in the range $-4.68-4.73 \times 10^{-7}$~cm$^3$/g (mass magnetic susceptibility) and
$-5.63-5.68 \times 10^{-6}$~cm$^3$/mole (molar magnetic susceptibility).
They are in very good correspondence with the experimental values: $-5.9 \times 10^{-6}$~cm$^3$/mole \cite{Hud}
and $-4.5 \times 10^{-7}$~cm$^3$/g~\cite{Her}, and previous LDA calculations of Mauri and Louie \cite{ML96}.

\acknowledgements

This work was supported by the National Natural Science Foundation of China (Grant No. 12274102) and
the Fundamental Research Funds for the Central Universities (Grant No. FRFCU5710053421, No. HIT.OCEF.2023031).

\appendix

\section{}
\label{appA}

Here we prove that the sum $\chi^{tot} = \chi^{dia} + \chi^{para}$ of the Langevin (Larmor) diamagnetic response $\chi^{dia}$ and
the Van Vleck paramagnetic response $\chi^{para}$ does not depend on the choice of the gauge origin [i.e. $\vec{R}_0$ in Eq.\ (\ref{i1a})], \cite{VV}.
The change of the vector potential $\vec{A}$ associated with $\vec{R}_0$ is in fact the degree of freedom, directly
related to the choice of the gauge. The gauge transformation for $\vec{A}$ can be written in the general form
$\vec{A} \rightarrow \vec{A} + \vec{\nabla} f$, where $f(\vec{r})$ is an arbitrary coordinate function.
In our case, Eq.\ (\ref{i1a}), the function $f$ is given by $f=-\vec{R}_0 \times \vec{H} \cdot \vec{r}/2$.
In Ref.\ \onlinecite{Lan} it is demonstrated that the gauge transformation does not alter essentially the solution,
resulting in the following transformation of the wave function: $\Psi \rightarrow \Psi\, \exp(-i e f(\vec{r})/\hbar c)$.
In particular, the electron energy $E(H)$ does not depend on the choice of $f(\vec{r})$ for any value of the magnetic field $H$.
This implies that the energy $E(H) = E_0 - \chi^{tot} H^2/2$,
taken in the second order of $H$, is also the same for any $f(\vec{r})$ and $\chi^{tot}$ does not depend on $\vec{R}_0$.
Since $\chi^{tot} = \chi^{dia} + \chi^{para}$, we have proved that the sum of Eq.\ (\ref{i1}) and Eq.\ (\ref{i3})
is independent of $\vec{R}_0$.

\section{}
\label{appB}

Here we consider the integrals ${\cal I}_{\,tj,\, \nu}$, where $\nu = x$, $y$, $z$, Eq.\ (\ref{ir2b}),
used in calculation of the matrix elements of the orbital momentum $L_H^{IR}$ in the interstitial region, Eq.\ (\ref{ir2a}).
Below we find them by means of Eq.\ (\ref{ir3a}). In the following we will condense notation
using ${K}_{\nu} \equiv {K}_{tj,\nu}$, ${\cal I}_{\nu} \equiv {\cal I}_{\,tj,\, \nu}$ and
limit ourselves to the case of interstitial region which has the inversion symmetry.
Then, starting with the standard expression
for the overlap integral $O_{tj}$ \cite{blapw,wien2k,Nik2}, we obtain
\begin{eqnarray}
   {\cal I}_{\nu} = {\cal I}_{\nu}^0 - i\, ({\cal I}_{\nu}^1 + {\cal I}_{\nu}^2 + {\cal I}_{\nu}^3) ,
\label{ab1}
\end{eqnarray}
where the ${\cal I}_{\nu}^0$ is connected with the plane wave integration over the whole unit cell, whereas ${\cal I}_{\nu}^n$ ($n=1,2,3$)
represent contributions from the MT-spheres, subtracted from ${\cal I}_{\nu}^0$.
Since the overlap over the unit cell is given by
\begin{eqnarray}
   O^{0} = \frac{2}{\vec{b_1}\vec{K}} \sin\! \left( \frac{\vec{b_1}\vec{K}}{2} \right)
   \frac{2}{\vec{b_2}\vec{K}} \sin \left( \frac{\vec{b_2}\vec{K}}{2} \right)
   \frac{2}{\vec{b_3}\vec{K}} \sin \left( \frac{\vec{b_3}\vec{K}}{2} \right) ,
 \nonumber   
\end{eqnarray}
where $\vec{b}_1$, $\vec{b}_2$, $\vec{b}_3$ are the corresponding basis vectors \cite{BC},
for ${\cal I}_{\nu}^0 \sim \partial O^{0}/ \partial K_{\nu}$, Eq.\ (\ref{ir3a}), we have nonzero contributions only if (1) $\vec{b_1}\vec{K} \neq 0$, $\vec{b_2}\vec{K} = \vec{b_3}\vec{K}=0$; or
(2) $\vec{b_2}\vec{K} \neq 0$, $\vec{b_1}\vec{K} = \vec{b_3}\vec{K}=0$; or (3) $\vec{b_3}\vec{K} \neq 0$, $\vec{b_1}\vec{K} = \vec{b_2}\vec{K}=0$.
In the first case, for example, we obtain
\begin{eqnarray}
   {\cal I}_{\nu}^0 = -i \frac{b_{1, \nu}}{\vec{b_1}\vec{K}} (-1)^m ,
\label{ab2}
\end{eqnarray}
where the integer $m=\vec{b_1}\vec{K} / 2 \pi$.
Further, in Eq.\ (\ref{ab1})
\begin{eqnarray}
   {\cal I}_{\nu}^1 &=& -\sum_{\alpha} C(\alpha) F'_{K,\, \nu}(\alpha, \vec{K})\, \frac{j_1(K R_{MT}^{\alpha})}{K} , \label{ab3} \\
   {\cal I}_{\nu}^2 &=& -\sum_{\alpha} C(\alpha)R_{MT}^{\alpha} F(\alpha, \vec{K})\, \frac{j'_1(K R_{MT}^{\alpha})}{K} \hat{K}_{\nu} , \nonumber \\
   {\cal I}_{\nu}^3 &=& \sum_{\alpha} C(\alpha) F(\alpha, \vec{K})\, \frac{j_1(K R_{MT}^{\alpha})}{K^2} \hat{K}_{\nu} .
 \nonumber 
\end{eqnarray}
Here $j_1(x)$ is the spherical Bessel function ($l=1$), $C(\alpha)=4 \pi (R_{MT}^{\alpha})^2/V$, $\hat{K}_{\nu}=K_{\nu}/K$, and
\begin{eqnarray}
   F(\alpha, \vec{K})=\sum_{\beta} \cos \left( \vec{K} \vec{R}_{\beta(\alpha)} \right) ,
 \nonumber 
\end{eqnarray}
where the summation on $\beta(\alpha)$ implies the summation over the equivalent atoms $\beta$ of the same atom type $\alpha$
centered at $\vec{R}_{\beta(\alpha)}$. Finally,
\begin{eqnarray}
   F'_{K,\, \nu} (\alpha, \vec{K})=-\sum_{\beta} {R}_{\beta(\alpha),\, \nu}\, \sin \left( \vec{K} \vec{R}_{\beta(\alpha)} \right) ,
 \nonumber 
\end{eqnarray}
where ${R}_{\beta(\alpha),\, \nu}$ is the $\nu$-th component of $\vec{R}_{\beta(\alpha)}$.
The same function appears in the contribution to $L_H^{MT,II}$ from displaced MT-spheres, Eq.\ (\ref{MT7}).
One can show that in general $F'_{K,\, \nu} (\alpha, \vec{K})= i\, {\cal R}_{\nu}$.

Note that all matrix elements for ${\cal I}_{\nu}$ are explicitly imaginary, Eqs.\ (\ref{ab1})--(\ref{ab3}).

\section{}
\label{appC}

Here we consider non-equivalent parts ${\cal N}_r$ required for the calculation of the Van Vleck summation for
selected directions of $H$.
Suppose we take $\vec{H}$ parallel to the [1,1,1]-axis (i.e. $\vec{n}_H=( 1,1,1)/ \sqrt{3}$) and
calculate contributions to $\chi^{para}$, Eq.\ (\ref{a2}), from different subparts ${\cal N}_r$.
In this particular case, both the Brillouin zone and the unit cell, Fig.~\ref{fig1}, have common symmetry operations
associated with the [1,1,1] three-fold rotation axis and some mirror planes (e.g. $x=y$).
As a result, all subsums to $\chi^{para}$ from first six parts ($n_0$, $n=1-6$) in Fig.\ \ref{fig2}
are equal. Among the remaining 18 parts (${\cal N}_r$, $n=1-6$, $r=1,2,3$) there are three groups with 6 equal contributions.
These findings are summarized in Table \ref{tab2}.
%
\begin{table}
\caption{
Partial contributions $\chi^{VV}({\cal N}_r)$
to the Van Vleck paramagnetic response $\chi^{para}=10.712$ for $\vec{H} \parallel [1,1,1]$
from various parts ${\cal N}_r$ of BZ, LDA calculations with $a_{latt}^{exp}$; $\chi$ -- volume values, in $10^{-7}$,
$W_0=48$, $N_{eq}$ is the number of equivalent parts in BZ,
$W$ is the total weight of $\chi^{VV}({\cal N}_r) \cdot W_0$ in $\chi^{para}$.
\label{tab2} }
\begin{ruledtabular}
\begin{tabular}{c  c  c  c }
 ${\cal N}_r$  &  $\chi^{VV}({\cal N}_r) \cdot W_0$ &  $N_{eq}$ &  $W$  \\
\tableline
  $1_0$ &  8.113 & 12 & 1/4 \\
  $1_1$ & 11.737 & 12 & 1/4 \\
  $1_2$ & 10.941 & 12 & 1/4 \\
  $1_3$ & 12.057 & 12 & 1/4
\end{tabular}
\end{ruledtabular}
\end{table}

If however we consider $\vec{H}$ parallel to the [1,1,-1]-axis (i.e. $\vec{n}_H=( 1,1,-1)/\sqrt{3}$)
then there will be no common three-fold rotation symmetry shared both the Brillouin zone and the unit cell.
In that case, in addition to inversion we have only the $x=y$ mirror plane, and the number of nonequivalent parts of the
Brillouin zone is increased from four, quoted in Table \ref{tab2}, to twelve, which are $1_r$, $3_r$ and $4_r$
($r=0,1,2,3$). The same nonequivalent parts of the Brillouin zone are involved if $\vec{H}$ is collinear to the [0,0,1]-axis,
Table \ref{tab3}, because in that case, as before, the only common point symmetry operation is the same $x=y$ mirror plane.

In general however, computation of all nonequivalent 24 contributions from the subparts of ${\cal N}_r$ is necessary for $\chi^{para}$.
In particular, this is required if $\vec{H}$ parallel to $[a,b,0]$ and $[-a,a,b]$ ($a \neq b$).

\section{}
\label{appD}


Here we reproduce and compare partial contributions to $\chi^{para}$ for $\vec{H} \parallel [0,0,1]$.
Earlier, we have analyzed the contributions $\chi^{VV}({\cal N}_r)$ from nonequivalent parts ${\cal N}_r$ of the Brillouin zone.
Now we will consider contributions to $\chi^{VV}({\cal N}_r)$ from various sources: the interstitial region, MT-sphere region,
and from the offset term $M_H^{MT,II}$, Eq.\ (\ref{MT3}).
Our results are given in Table \ref{tab3}.
%
\begin{table}
\caption{
Contributions $\chi'=\chi^{VV} \cdot W_0$ to the Van Vleck paramagnetic response $\chi^{para}=23.136$ for $\vec{H} \parallel [0,0,1]$
from the interstitial region (IR), MT-sphere region (MT) and the offset part with $M_H^{MT,II}$, Eq.\ (\ref{MT3}),
for various parts ${\cal N}_r$ of BZ, LDA calculations with $a_{latt}^{exp}$; all $\chi$ are volume values,
in units $10^{-7}$, $W_0=48$; see text for details.
\label{tab3} }
\begin{ruledtabular}
\begin{tabular}{c  c  c  c  c}
 ${\cal N}_r$  &  $\chi'$, IR & $\chi'$, MT & $\chi'$, offset & $\chi'$, total  \\
\tableline
  $1_0$ & 14.836 & 12.951 & 7.480 & 22.010 \\
  $1_1$ & 15.020 & 12.708 & 7.148 & 22.153 \\
  $1_2$ & 15.173 & 13.265 & 7.480 & 22.577 \\
  $1_3$ & 15.099 & 12.845 & 7.148 & 22.547 \\
  $3_0$ & 15.385 & 14.717 & 7.450 & 26.256 \\
  $3_1$ & 15.000 & 14.459 & 7.230 & 25.500 \\
  $3_2$ & 15.068 & 14.582 & 7.450 & 25.153 \\
  $3_3$ & 15.255 & 14.399 & 7.230 & 25.705 \\
  $4_0$ & 14.927 & 12.055 & 7.387 & 20.372 \\
  $4_1$ & 15.201 & 12.404 & 7.240 & 21.607 \\
  $4_2$ & 15.521 & 12.872 & 7.387 & 22.500 \\
  $4_3$ & 15.229 & 12.228 & 7.240 & 21.248 \\
\end{tabular}
\end{ruledtabular}
\end{table}

It is worth noting that we obtain contributions from these sources to the total magnetic moment $M_H$ in matrix form, Eq.\ (\ref{a3}),
using LAPW basis functions $\phi_j$, Eq.\ (\ref{m1}).
At that level we have the property of additivity, i.e. $\langle t | M_H | j \rangle = \langle t | M_H^{IR} | j \rangle + \langle t | M_H^{MT} | j \rangle$.
However, then we transform the matrices to the basis of band eigenstates, after which the magnetic moments are squared and substituted in Eq.\ (\ref{a2}).
The last procedure obviously destroys the initial additivity.
Therefore, in obtaining a particular contribution in Table \ref{tab3} the other terms were artificially put to zero.
The MT contribution $M_H^{MT}$ includes the offset contributions $M_H^{MT,II} = \mu_B L_H^{MT,II}$, Eq.\ (\ref{MT2}), and Eq.\ (\ref{MT3}).
The column with the offset contributions in Table \ref{tab3} was calculated when $M_H = M_H^{MT,II}$.
Inspection of the data of Table \ref{tab3} indicates that the contributions from the IR and MT regions are almost equal for all ${\cal N}_r$,
with the IR contribution being somewhat larger. The offset contribution, on the other hand, is typically smaller -- approximately 1/3 of the total value of $\chi^{VV}$.
Nevertheless, it accounts for more than half of the MT-contribution.

All partial contributions $\chi$ in Tables \ref{tab2} and \ref{tab3} are enhanced by the factor $W_0=48$. Thus, they are scaled to
the whole Brillouin zone as if all parts were equivalent to the chosen ${\cal N}_r$.
In reality, for every row (${\cal N}_r$) in Table \ref{tab3} there are only four equivalent parts ($N_{eq}$) in BZ with the same value.
Therefore, the weight of every part in Table \ref{tab3} is $W=1/12$, and to obtain the true contribution from the whole BZ one has to average the values in columns.
That results in $\chi^{VV}(IR)$=15.143 for the interstitial region, $\chi^{VV}(MT)$=13.290  for the MT region, the offset value is $\chi^{VV}(MT,II)$=7.323,
and the total Van Vleck contributions is $\chi^{VV}$=23.136,
in units of $10^{-7}$ (volume values).



\begin{thebibliography}{99}

\bibitem{spin}
D. Awschalom, M. Flatt\'e,
Challenges for semiconductor spintronics.
Nature Phys. {\bf 3}, 153 (2007).

\bibitem{Barry}
J. F. Barry, J. M. Schloss, E. Bauch, M. J. Turner, C. A. Hart, L. M. Pham, et al.
Sensitivity optimization for NV-diamond magnetometry.    
Rev. Mod. Phys. {\bf 92}, 015004 (2020).

\bibitem{diam-22}
N. Nunn, M. D. Torelli, A. Ajoy, A. I. Smirnov, and O. Shenderova,
Beauty beyond the Eye: Color Centers in Diamond Particles
for Imaging and Quantum Sensing Applications,
Reviews and Advances in Chemistry, {\bf 12}, 1 (2022).

\bibitem{King2}
J.P. King, K. Jeong, C.C. Vassiliou, C.S. Shin, R.H. Page, C.E. Avalos, H-J. Wang, A. Pines,
Room-temperature in situ nuclear spin hyperpolarization from optically pumped nitrogen vacancy centres in diamond,
Nat. Commun. {\bf 6}, 8965 (2015).

\bibitem{King}
J.P. King, P.J. Coles, J.A. Reimer,
Optical polarization of $^{13}$C nuclei in diamond through nitrogen vacancy centers,
Phys. Rev. B {\bf 81}, 073201 (2010).

\bibitem{C13-rev}
J. Henshaw, D. Pagliero, P. R. Zangara, M. B. Franzoni, A. Ajoy, R. H. Acosta,
J. A. Reimer, A. Pines, and C. A. Meriles,
Carbon-13 dynamic nuclear polarization in diamond via a microwave-free integrated cross effect,
Proc. Natl. Acad. Sci. USA 116, 18334 (2019).

\bibitem{Mul}
C. M\"uller, X. Kong, J.-M. Cai, K. Melentijevi\'c, A. Stacey, M. Markham, et al.
Nuclear magnetic resonance spectroscopy with single spin sensitivity.
Nat. Commun. {\bf 5}, 4703 (2014).

\bibitem{Bala}
G. Balasubramanian, et al.
Nanoscale imaging magnetometry with diamond spins under ambient conditions.
Nature {\bf 455}, 648 (2008).

\bibitem{Maze}
J. R. Maze, et al.
Nanoscale magnetic sensing with an individual electronic spin in diamond.
Nature {\bf 455}, 644–647 (2008).

\bibitem{Zhao}
N. Zhao, J.-L. Hu, S.-W. Ho, J. T. K. Wan, and R.-B. Liu,
Atomic-scale magnetometry of distant nuclear spin clusters via nitrogen-vacancy spin in diamond.
Nat. Nanotech. {\bf 6}, 242 (2011).

\bibitem{McGui}
L. P. McGuinness, et al.
Quantum measurement and orientation tracking of fluorescent nanodiamonds inside living cells.
Nat. Nanotechnol. {\bf 6}, 358 (2011).

\bibitem{Buck}
B. B. Buckley, G. D. Fuchs, L. C. Bassett, and D. D. Awschalom,
Spin-light coherence for single-spin measurement and control in diamond.
Science {\bf 330}, 1212 (2010).

\bibitem{Pez}
S. Pezzagna, and J. Meijer,
Quantum computer based on color centers in diamond.
Appl. Phys. Rev. {\bf 8}, 011308 (2021).

\bibitem{Stev}
G. Stevanato, J. T. Hill-Cousin, P. H{\aa}kansson, S. S. Roy, L. J. Brown, R. C. Brown, et al.
A nuclear singlet lifetime of more than one hour in room-temperature solution.
Angew. Chem. Int. Ed. {\bf 54}, 3740 (2015).

\bibitem{Shagi}
F. Shagieva, S. Zaiser, P. Neumann, D.B.R. Dasari, R. St\"ohr, A. Denisenko, R. Reuter, C.A. Meriles, J. Wrachtrup,
Microwave assisted cross-polarization of nuclear spin ensembles from optically pumped nitrogen-vacancy centers in diamond,
Nano Lett. {\bf 18}, 3731 (2018).

\bibitem{Green}
B.L. Green, B.G. Breeze, G.J. Rees, J.V. Hanna, J.-P. Chou, V. Iv\'ady, A. Gali, M.E. Newton,
All-optical hyperpolarization of electron and nuclear spins in diamond,
Phys. Rev. B {\bf 96}, 054101 (2017).

\bibitem{Pagli}
D. Pagliero, K.R. Koteswara Rao, P.R. Zangara, S. Dhomkar, H.H. Wong, A. Abril, N. Aslam, A. Parker, J. King, C.E. Avalos, A. Ajoy, J. Wrachtrup, A. Pines, C.A. Meriles,
Multispin-assisted optical pumping of bulk $^{13}$C nuclear spin polarization in diamond,
Phys. Rev. B {\bf 97}, 024422 (2018).

\bibitem{Niz}
A. P. Nizovtsev, S.Ya. Kilin, A. L. Pushkarchuk, V. A. Pushkarchuk, S. A. Kuten, H. Zhikol, et al.
Non-flipping $^{13}$C spins near an NV center in diamond:
hyperfine and spatial characteristics by density functional theory simulation of the C510 [NV]H252 cluster.
New J. Phys. {\bf 20}, 023022 (2018).

\bibitem{Hud}
S. Hudgens, M. Kastner, and H. Fritzsche,
Diamagnetic Susceptibility of Tetrahedral Semiconductors,
Phys. Rev. Lett. {\bf 33}, 1552 (1974).

\bibitem{Her}
J. Heremans, C.H. Olk, and D.T. Morelli,
Magnetic susceptibility of carbon structures,
Phys. Rev. B {\bf 49}, 15122 (1994).

\bibitem{Nik-book}
A.V. Nikolaev and B. Verberck,
Diamagnetism of Diamond and Graphite,
in {\it Carbon based magnetism}, Ed. T. Makarova and F. Palacio, Elsevier (2006), p.\ 245.

\bibitem{Sukh75}
V. P. Sukhatme and P. A. Wolff,
Chemical-Bond Appproach to the Magnetic Susceptibility of Tetrahedral Semiconductors,
Phys. Rev. Lett. {\bf 35}, 1369 (1975).

\bibitem{Chadi75}
D. J. Chadi, R. M. White, and W. A. Harrison,
Theory of the Magnetic Suusceptibility of Tetrahedral Semiconductors,
Phys. Rev. Lett. {\bf 35}, 1372 (1975).

\bibitem{Nik-VV}
A.V. Nikolaev, M.Ye. Zhuravlev, L.L. Tao,
Ab initio based study of the diamagnetism of diamond, silicon and germanium,
J. Magn. Magn. Mater. {\bf 588}, 171394 (2023).

\bibitem{ML96}
F. Mauri and S. G. Louie,
Magnetic Susceptibility of Insulators from First Principles,
Phys. Rev. Lett. {\bf 76}, 4246 (1996).

\bibitem{Gregor99}
T. Gregor, F. Mauri, and R. Car,
A comparison of methods for the calculation of NMR chemical shifts,
J. Phys. Chem. {\bf 111}, 1815 (1999).

\bibitem{MPL}
F. Mauri, B. G. Pfrommer, and S. G. Louie,
Ab Initio Theory of NMR Chemical Shifts in Solids and Liquids,
Phys. Rev. Lett. {\bf 77}, 5300 (1996).

\bibitem{Pickard01}
C. J. Pickard and F. Mauri,
All-electron magnetic response with pseudopotentials: NMR chemical shifts,
Phys. Rev. B {\bf 63}, 245101 (2001).

\bibitem{Yates07}
J. R. Yates, C. J. Pickard, and F. Mauri,
Calculation of NMR chemical shifts for extended systems using ultrasoft pseudopotentials,
Phys. Rev. B {\bf 76}, 024401 (2007).

\bibitem{Laskow12}
R. Laskowski and P. Blaha,
Calculations of NMR chemical shifts with APW-based methods,
Phys. Rev. B {\bf 85}, 035132 (2012).

\bibitem{Laskow14}
R. Laskowski and P. Blaha,
Calculating NMR chemical shifts using the augmented plane-wave method,
Phys. Rev. B {\bf 89}, 014402 (2014).

\bibitem{VV}
J. H. Van Vleck. {\it The Theory of Electric and Magnetic Susceptibilities}
(Oxford University Press, London 1932), p. 276.

\bibitem{AM}
N.W. Ashcroft, and N.D. Mermin, (1976) {\it Solid State Physics}. Brooks/Cole Cengage Learning (1976).

\bibitem{Lan1}
L. D. Landau and E. M. Lifshitz, {\it Statistical Physics} (Pergamon, Bristol, 1995), Vol.~5.

\bibitem{Kit}
Ch. Kittel, {\it Introduction to solid state physics} -- 8th ed. John Wiley \& Sons (2005).

\bibitem{Nik1}
A. V. Nikolaev,
Landau diamagnetic response in metals as a Fermi surface effect,
Phys. Rev. B, {\bf 98}, 224417 (2018).

\bibitem{blapw}
D.J. Singh, L. Nordstr\"{o}m, {\it Planewaves, Pseudopotentials, and the LAPW Method}, 2nd ed. (Springer, New York, 2006).

\bibitem{wien2k} P. Blaha, K. Schwarz, G. Madsen, D. Kvasnicka and J. Luitz,
J. Luitz, WIEN2K: {\it An Augmented Plane Wave plus Local
Orbitals Program for Calculating Crystal Properties}
(Vienna University of Technology, Austria, 2001).

\bibitem{Nik2}
A. V. Nikolaev, D. Lamoen, and B. Partoens,
Extension of the basis set of linearized augmented plane wave (LAPW)
method by using supplemented tight binding basis functions,
J. Chem. Phys. {\bf 145}, 014101 (2016).

\bibitem{Lan}
L. D. Landau and E. M. Lifshitz, {\it Quantum Mechanics - Non-relativistic theory} (Pergamon, Bristol, 1995), Vol.~3.

\bibitem{BC}
 C.~J. Bradley and A.~P. Cracknell,
 {\it The Mathematical Theory of Symmetry in Solids},
 (Clarendon, Oxford, 1972).

\bibitem{tet}
G. Lehmann and M. Taut,
On the numerical calculation of the density of states and related properties,
Phys. Status Solidi B {\bf 54}, 469 (1972).

\bibitem{PBE} J. P. Perdew, K. Burke, and M. Ernzerhof,
Generalized Gradient Approximation Made Simple,
Phys. Rev. Lett. {\bf 77}, 3865 (1996).

\bibitem{Dir}
P. A. M. Dirac,
Note on Exchange Phenomena in the Thomas Atom,
Proc. Camb. Philos. Soc. {\bf 26}, 376 (1930).

\bibitem{PWcorr}
J. P. Perdew and Y. Wang,
Accurate and simple analytic representation of the electron-gas correlation energy,
Phys. Rev. B {\bf 45}, 13244 (1992).

\bibitem{Nik3}
A. V. Nikolaev, I. T. Zuraeva, G. V. Ionova,
Spin-polarization and spin-orbit interactions in the LAPW method: application in description of 3d metals,
and B. V. Andreev, Fizika Tverdogo Tela {\bf 35}, 414 (1993).
[translation: Phys. Solid State {\bf 35}, 213 (1993)].

\bibitem{Varsh}
D. A. Varshalovich, A. N. Moskalev, and V. K. Khersonskii
{\it Quantum Theory of Angular Momentum and its applications}, vol. 1, Moscow, Fizmatlit (2017).

\bibitem{Leb1}
V. I. Lebedev,
Values of the nodes and weights of ninth to seventeenth order gauss-markov quadrature formulae invariant under the octahedron group with inversion,
USSR Computational Mathematics and Mathematical Physics, {\bf 15}, 44 (1975).

\bibitem{Leb2}
V.I. Lebedev, and D.N. Laikov,
A quadrature formula for the sphere of the 131st algebraic order of accuracy,
Doklady Mathematics, {\bf 59}, 477, (1999).

\end{thebibliography}
\end{document}